\documentclass[fleqn,usenatbib]{mnras}

\usepackage{newtxtext,newtxmath}

\usepackage[T1]{fontenc}
\usepackage{ae,aecompl}

\usepackage{graphicx}	
\usepackage{amsmath}	
\usepackage{amssymb}	
\usepackage{color, colortbl}
\usepackage{afterpage}
\usepackage{pdflscape}
\definecolor{linkcolor}{rgb}{0,0,0.25}
\hypersetup{
  colorlinks=true,        
  linkcolor=linkcolor,    
  citecolor=linkcolor,    
  filecolor=linkcolor,    
  urlcolor=linkcolor      
}

\makeatletter
\renewcommand{\@printed}{}
\makeatother

\newcommand{\figurename}{Figure}
\newcommand{\tablename}{Table}
\newcommand{\eqnname}{Equation}

\definecolor{darkgreen}{rgb}{0.0, 0.5, 0.0}

\definecolor{darkblue}{rgb}{0.0, 0., 0.7}

\definecolor{purple}{rgb}{0.7, 0., 0.7}

\definecolor{Gray}{gray}{0.75}

\newcommand{\gaia}{\emph{Gaia}}

\title[Red clump \gaia\ DR2]{The \gaia\ DR2 parallax zero point: Hierarchical modeling of red clump stars}

\author[Chan \& Bovy]{
Victor C. Chan\thanks{E-mail: chan@astro.utoronto.ca}\&
Jo Bovy
\\
Department of Astronomy and Astrophysics, University of Toronto, 50 St. George Street, Toronto, Ontario, M5S 3H4, Canada
}

\date{Accepted 24 February 2020. Received 2020 February 3; in original form 2019 September 29}

\pubyear{2020}

\begin{document}
\label{firstpage}
\pagerange{\pageref{firstpage}--\pageref{lastpage}}
\maketitle

\begin{abstract}
The systematic offset of \gaia\ parallaxes has been widely reported with \gaia's second data release, and it is expected to persist in future \gaia\ data. In order to use \gaia\ parallaxes to infer distances to high precision, we develop a hierarchical probabilistic model to determine the \gaia\ parallax zero point offset along with the calibration of an empirical model for luminosity of red clump stars by combining astrometric and photometric measurements. Using a cross-matched sample of red clump stars from the Apache Point Observatory Galactic Evolution Experiment (APOGEE) and \gaia\ Data Release 2 (DR2), we report the parallax zero point offset in DR2 to be $ \varpi_0 = -48 \pm 1~\mu\text{as} $. We infer the red clump absolute magnitude to be $ -1.622 \pm 0.004 $ in $ K_s $, $ 0.435 \pm 0.004 $ in $ G $, $ -1.019 \pm 0.004 $ in $ J $, and $ -1.516 \pm 0.004 $ in $ H $. The intrinsic scatter of the red clump is $ \sim 0.09 $ mag in $ J $, $ H $ and $ K_s $, or $ \sim 0.12 $ mag in $ G $. We tailor our models to accommodate more complex analyses such as investigating the variations of the parallax zero point with each source's observed magnitude, observed colour, and sky position. In particular, we find fluctuations of the zero point across the sky to be of order or less than a few 10s of $ \mu\text{as} $.
\end{abstract}




\section{Introduction}

The \gaia\ satellite is tasked with performing the world's largest simultaneous astrometric, photometric, and spectroscopic survey \citep{GaiaMission}. With its second data release (DR2) in April 2018, the survey accrued astrometric and photometric measurements for over one billion sources down to a magnitude of $ G \lesssim 21 $ \citep{GaiaDR2Summary}. The \gaia\ satellite measures absolute parallaxes by comparing positions of stars in two fields of view (FOVs) widely separated by the ``basic" angle $ \Gamma_\text{f} = 106.5^{\circ} $ along the plane of its scanning motion. For a source passing through the center of one of the FOVs, its position is described by $ \phi = \pm \Gamma_\text{f}/2 $ with respect to the axis bisecting the \textit{formal} basic angle in the same plane. An analytical solution shows that perturbations to such a source's observed parallax ($ \varpi $) is degenerate with perturbations to the true basic angle $ \Gamma $ (Equation 15; \citealt{ZPOrigin}), repeated here
\begin{equation}\label{eq:BasicAngleToPlx}
	\delta\Gamma = \zeta \sin(\phi) \delta\varpi = \zeta \sin(\Gamma_\text{f}/2) \delta\varpi,
\end{equation}
where $ \zeta $ is a function of the satellite's orientation with respect to the Solar System barycenter. In other words, oscillations in the instrument's basic angle are degenerate with an absolute perturbation in the observed parallax. This effect is largely corrected by the on-board basic angle monitor, but the scanning motions of the spacecraft leave residual contributions to the parallax zero point that are difficult to model.

A similar effect is also expected to arise from the use of distinct calibration units for sources of different apparent magnitudes \citep{PhotometryProcess}. In addition, \gaia's astrometric solution relies on the position of each source's centroid on the CCD, which is affected by the brightness of the source. There were two methods of determining the positions of centroids: both dim and bright sources ($ G \gtrsim 13 $ and $ G < 13 $ respectively) have the positions of their centroids measured as they cross a fiducial line on the CCD, and bright sources $ G > 13 $ also have their orthogonal positions determined with the entire column traced out by the centroids as they drift across the detector \citep{GaiaDR2Astrometry}. This extra dimension of position measurement could leave different systematics within the astrometric measurements. In addition, the astrometric calibration used in DR2 makes use of an effective wavenumber determined using mean integrated blue ($ G_{BP} $) and red ($ G_{RP} $) photometric magnitudes. The unique observed colour of each source is expected to contribute to fluctuations to the astrometric solution, equivalent to a parallax zero point dependence on observed colour. Finally, the parallax zero point has also been shown to vary on large scales across the sky by mapping the observed parallaxes of quasars in the DR2 sample \citep{GaiaDR2CatValid}.

In general, \gaia\ parallaxes have been reported to be too small. The parallax zero point offset in DR2 has been reported by the \gaia\ collaboration to be $ \varpi_0 = -29 \pm 1~\mu\text{as} $, measured using the parallaxes of quasars in the DR2 sample \citep{GaiaDR2Astrometry}. This is further supported when they compare DR2 parallaxes of various globular clusters to those found in literature ($ \varpi_0 \approx -25~\mu\text{as} $), but a slightly different result was found using dwarf spheroidal galaxies ($ \varpi_0 \approx -49~\mu\text{as} $) \citep{GCAstrometry}. \citet{RiessDR2} found $ \varpi_0 = -46 \pm 13~\mu\text{as} $ by combining \gaia\ parallaxes with Hubble Space Telescope photometry of Milky Way Cepheids. \citet{EclBin1} report $ \varpi_0 = -82 \pm 33~\mu\text{as} $, and \citet{EclBin2} determine a zero point of $ \varpi_0 = -31 \pm 11~\mu\text{as} $ from analyses comparing existing measurements of parallax for eclipsing binary stars with \gaia\ parallaxes. \citet{AstSeisRGB} report a modelled $ \varpi_0 = -52.8 \pm 2.4 $ (random) $ \pm 8.6 $ (systematic) $ - (150.7 \pm 22.7)(\nu_\text{eff} - 1.5) - (4.21 \pm 0.77)(G - 12.2) \mu\text{as} $ with significant dependence on the effective wavenumber and observed magnitude using asteroseismology of red giant branch stars. Similarly, \citet{AstSeisDwarf} report a parallax zero point of $ \varpi_0 = -35 \pm 16~\mu\text{as} $ using the asteroseismology of dwarf stars, and \citet{HallAstSeis} use the asteroseismology of red clump stars to determine a mean parallax zero point of $ \varpi_0 = -41 \pm 10~\mu\text{as} $, with individual estimates of $ \varpi_0 = -38 \pm 13~\mu\text{as} $ in $ K_s $ and $ \varpi_0 = -42 \pm 13~\mu\text{as} $ in $ G $. Using deep learning of spectro-photometric distances, \citet{AstroNNZP} determine a modeled constant zero point of $ \varpi_0 = -52.3 \pm 2.0~\mu\text{as} $ and they present quadratic parameterizations of the zero point's dependences on $ G $, observed colour, and effective temperature.

In this paper, we describe hierarchical Bayesian models inspired by \citet{SesarRRLyrae} and \citet{HawkinsRC} to simultaneously estimate the \gaia\ parallax zero point and constrain an empirical relation for the luminosity of red clump stars. We outline the red clump sample and the data used in these analyses in \S\ref{sec:Data}. The hierarchical model and several add-ons for additional detailed analysis are described in \S\ref{sec:Methods}. Results of each model/analysis are presented in \S\ref{sec:Results}, with inferred parameters collected in \tablename~\ref{tab:results}. We discuss the internal consistencies between each model as well as compare our measurements with other reports in \S\ref{sec:Discussion}. Finally, we conclude with a summary in \S\ref{sec:Conclusion}.

\section{Data}\label{sec:Data}

\subsection{Red-clump sample}

As a part of SDSS-III/IV \citep{SDSSIII, SDSSIV}, the Apache Point Observatory Galactic Evolution Experiment (APOGEE; \citealt{APOGEE}) is a spectroscopic survey in the near infrared. Observing in the infrared is a major advantage for its data, as this allows for measurements that are less affected by dust extinction when compared to optical measurements. The survey data has been pre-processed and is publicly available online as part of the SDSS Data Release 14 (DR14; \citealt{SDSSAnalysis, ASPCAP, SDSSDR14}). This data set includes detailed measurements of each source's chemical abundances (with $S/N$ > 100) as well as the stellar parameters $T_\mathrm{eff}$ and $\log g$ due to its spectroscopic resolution of $R \approx 22~500$ \citep{2019PASP..131e5001W}. We make use of the Two Micron All Sky Survey (2MASS) $J$ , $H$ , and $K_s$ photometry \citep{2MASS}. The 2MASS photometry has been previously corrected for reddening with the Rayleigh-Jeans Colour Excess method (Equation 1; \citealt{RJCE}), yielding measured extinction values $A_J$, $A_H$, and $A_K$ for each object.

We also make use of inferred values for $T_\mathrm{eff}$, $\log g$, as well as the chemical abundance data, specifically [Fe/H], [O/Fe], [Mg/Fe], [Si/Fe], [S/Fe], and [Ca/Fe]. The abundances for $Z = \{\textrm{O, Mg, Si, S, Ca}\}$ are used to construct a value of [$\alpha$/Fe] by calculating
\begin{equation}\label{eq:alphaFe}
	[\alpha/\text{Fe}] = \frac{\sum_{z \in Z} w_z [z\text{/Fe}]}{\sum_{z \in Z} w_z},
\end{equation}
where $w=1$ or $w=0$ depending on whether the measurement of [Z/Fe] exists respectively. The abundance measurements upon which $ [\alpha/\text{Fe}] $ is based are not taken from SDSS DR14, but they are instead products of an artificial neural network \verb|astroNN| trained on APOGEE spectra from DR14 \citep{astroNN}. Typical sources in \verb|astroNN| have inferences of chemical abundances to $ \approx 0.03~\text{dex} $, $ T_\mathrm{eff}$ to $ \approx 30~\text{K} $, and $ \log g $ to $ \approx 0.05~\text{dex} $.

We use only a sub-sample of the APOGEE data set that has been classified as red-clump stars by \citet{BovyRC}. The red-clump stars were identified with strict cuts first in effective-temperature--surface-gravity--metallicity space followed by further cuts in colour--surface-gravity--metallicity space. This was done to minimize contamination from other red giant branch stars within the population.

The \citet{BovyRC} cuts leave a small amount of contamination $(\lesssim 10\%)$ by non red-clump stars, which could be identified with better measurements of $\log g$. In order to further purify the red-clump sample, we apply cuts to sources in the sample for which $ (\log g)_\text{APOGEE} - (\log g)_\texttt{astroNN} < -0.2 $. As this quality cut serves to remove suspected contamination sources in the sample, it should improve our calibrations of the red clump luminosity. We do not expect there to be any bias induced on statistical inferences. We provide a brief discussion on the importance of an accurate red clump luminosity calibration in \S\ref{sec:DiscussionConsistency}.

\subsection{\emph{Gaia} DR2 data}

With the advent of DR2, the \gaia\ mission currently has measurements of sky position and photometry in its $G$-band (330--1050 nm) for nearly 1.7 billion sources \citep{GaiaDR2Summary}. Over 1.3 billion of the sources in DR2 also have measurements of parallax ($\varpi$) and proper motion. The current reported limiting magnitude in DR2 is $G \approx 21$, and the data set is reported to be complete over $3 \lesssim G \lesssim 17$.

The \gaia\ mission only reports the 5-parameter astrometric data (position, parallax, and proper motion) for sources that satisfy three requirements: (1) they must be brighter than the limiting magnitude ($G < 21$), (2) each source must have been observed on at least six occasions that are separated by at least four days, and (3) the astrometric parameters must be measured to within a magnitude-dependent uncertainty. The full details of the astrometric solution are discussed by \citet{GaiaDR2Astrometry}.

We cross-match the \gaia\ DR2 sample with the red-clump sample to obtain the $G$ magnitudes, $G_{BP}$ (330--680 nm; also referred to as $ BP $) and $G_{RP}$ (630--1050 nm; also referred to as $ RP $) magnitudes, as well as parallaxes and their uncertainties. The \gaia\ data set includes measurements of negative parallaxes, which can still contain useful information when combined with their uncertainties. An exception must be made for unphysical measurements of negative parallaxes that are too confident because these must be outliers, so we make further cuts on the quality of the parallax measurements by removing any sources with a measured $\varpi/\sigma_\varpi < -3$, or $\varpi = 0$ from our sample.

\begin{figure}
	\centering
	\includegraphics[width=\hsize]{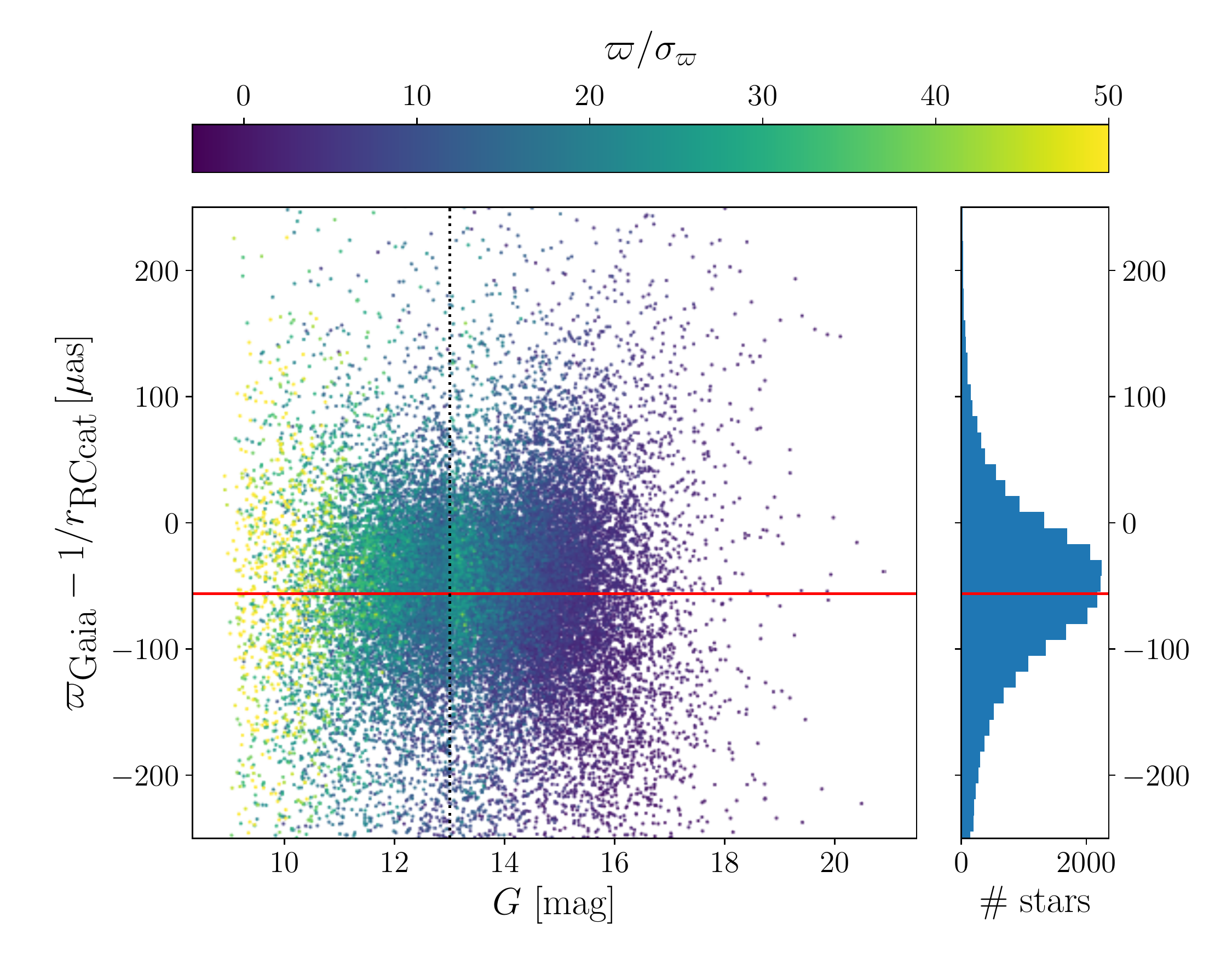}
	\caption{The \gaia\ parallax offset assuming the distances to each star in the red clump catalogue are correct. The red line indicates the median of the sample at $-56.2~\mu\text{as}$. The dotted black line is located at $ G = 13 $ mag, where the astrometric solution pipeline is reported to change.}
	\label{fig:RCOffset}
\end{figure}

Additionally, the \gaia\ collaboration has advised that the $ G $ magnitudes be corrected according to the following procedure\footnote{\url{https://www.cosmos.esa.int/web/gaia/dr2-known-issues}}
\begin{equation}\label{eq:Gcorr}
	G_\text{corr} = 
	\begin{cases}
	-0.0473 + 1.164 G - 0.0468 G^2 + 0.0035 G^3,~ 2 < G \leq 6 \\
	G - 0.0032(G-6),~ 6 < G \leq 16 \\
	G - 0.032,~ G > 16.
	\end{cases}
\end{equation}
This correction is applied only when the $ G $ magnitudes are used in the context of using $ G $ to determine the red clump luminosity through the distance modulus (further elaborated upon in \S\ref{sec:Methods}). Our final data set contains 27,934 red clump stars with values for $\mathcal{D} = \big\{\varpi, \sigma_\varpi, G, G_{BP}, G_{RP}, J, H, K_s, A_J, A_H, A_{K_s}, T_\text{eff}, [\text{Fe/H}], [\alpha\text{/Fe}]\big\}$. The data set is shown in \figurename~\ref{fig:RCOffset}, where observed DR2 parallax is compared to distances from the APOGEE red clump catalogue as a function of $ G $.

\subsection{\gaia\ Extinction Model}\label{sec:ExtinctionModel}

As mentioned previously, the 2MASS photometry from the red clump catalog ($J$, $H$, and $K_s$ bands) have been corrected for extinction using the Rayleigh-Jeans colour excess (RJCE) method, so each star has associated $A_J$, $A_H$, and $A_K$ extinction values that have been implicitly included in their respective apparent magnitudes as $m_0 = m - A_m$. The \gaia\ $G$ band has \emph{not} been extinction corrected, so we use the following procedure to estimate the $G$ band extinction coefficient.

\begin{figure}
    \centering
    \includegraphics[width=\hsize]{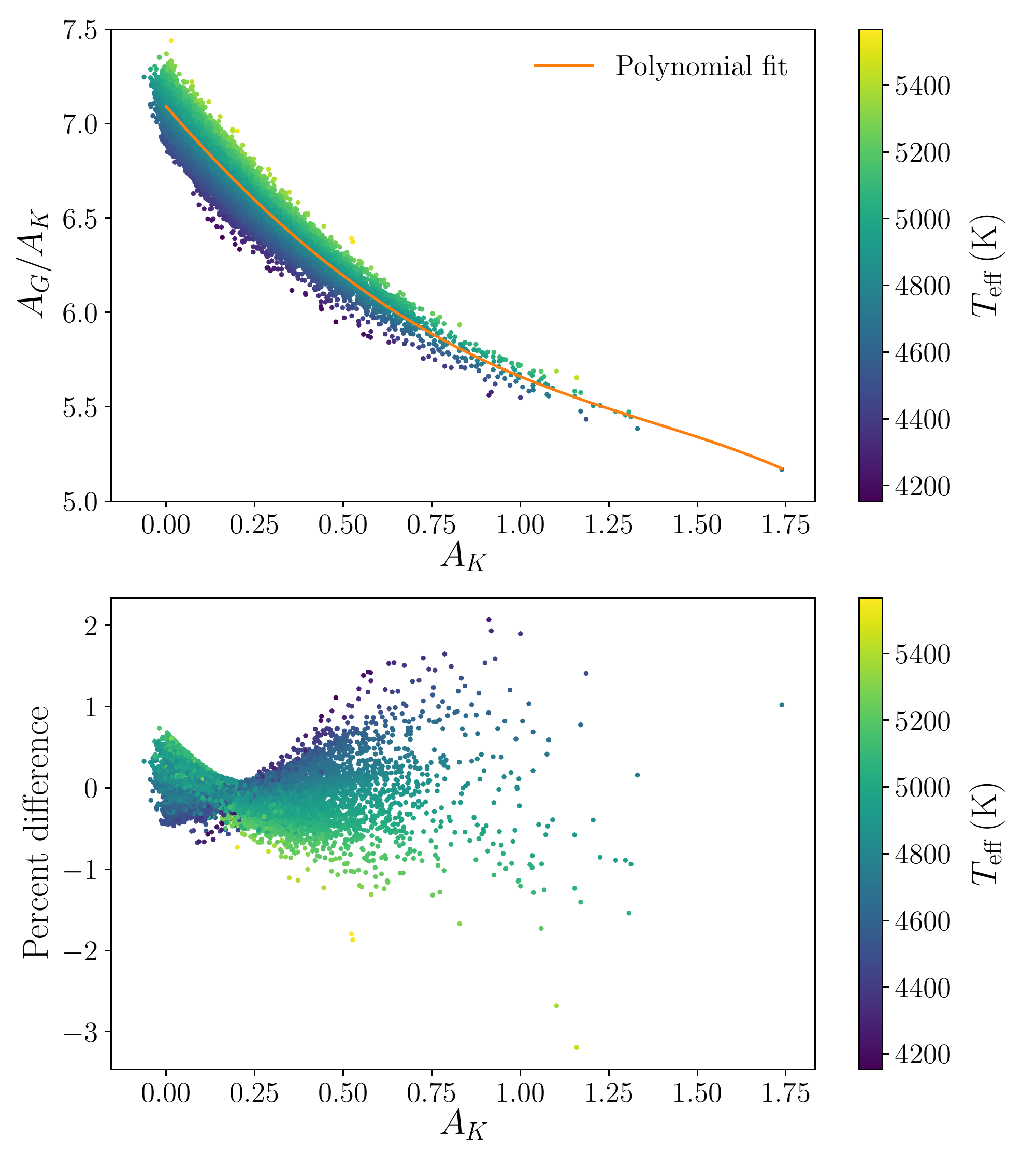}
    \caption{The stellar extinction ratios $ A_G/A_K $ for the red clump sample. \textit{Top:} The sample of extinction ratios using the steps from \S\ref{sec:ExtinctionModel}, along with the polynomial fit described by \eqnname~\eqref{eq:AGAKfit}. \textit{Bottom:} The residuals from the polynomial fit. \eqnname~\eqref{eq:AGAKfit} describes the sample of modeled extinction ratios to within a few percent.}
    \label{fig:AGAKs}
\end{figure}

For each star in the red clump sample, we compute synthetic $ G $ and $ K_s $ band photometry using the \texttt{pystellibs}\footnote{\url{https://github.com/mfouesneau/pystellibs}}, \texttt{pyphot}\footnote{\url{https://github.com/mfouesneau/pyphot}}, and \texttt{pyextinction}\footnote{\url{https://github.com/mfouesneau/pyextinction}} tools, and we determine the $A_G/A_K$ extinction ratios with the following steps:
\begin{enumerate}
	\item We generate initial synthetic stellar spectra (denoted $ F_0(\lambda) $) with the \citet{KuruczATLAS9}/\citet{KuruczStellarModels} stellar model library in \texttt{pystellibs} for each source using their respective $T_\textrm{eff}$, $\log g$, and $Z$.
	
	\item The extinction-free $ K_s $ band magnitude (denoted $K_0$) is computed with \texttt{pyphot} for each stellar spectrum $ F_0(\lambda) $ using the 2MASS $ K_s $ passband \citep{2MASSKsPassband}.
	
	\item Assuming a value of $ A_V $, we use the \citet{FitzpatrickExtinction} extinction law $A(\lambda) / A_V$ computed using the \texttt{pyextinction} tool, redden each stellar spectrum with $ A(\lambda) = A_V \times A(\lambda)/A_V $, yielding $ F_r(\lambda) = F_0(\lambda) \times 10^{-0.4 A(\lambda)} $. We then require knowledge of $ A_V $ for each star in order to apply the correct extinction to each stellar spectrum.
	
	\item The extinguished $ K_s $ band magnitude (denoted $ K $) is computed with \texttt{pyphot} for each reddened spectrum $ F_r(\lambda) $ using the 2MASS $ K_s $ passband.
	
	\item By repeated application of the previous two steps, we solve for each $A_V$ value that matches the given $A_K$ value obtained from the RJCE method. We then redden each stellar spectrum again using this value of $A_V$, yielding $ F_R(\lambda) $.
	
	\item The extinction-free $ G_0 $ magnitude is computed with \texttt{pyphot} for each un-reddened stellar spectrum $ F_0(\lambda) $ using the \citet{MAWPassband} revised $ G $ passband. Similarly, we compute each extinguished $ G $ from the reddened spectrum $ F_R(\lambda) $.
	
	\item The extinction ratio for each star is then produced as $ A_G/A_K = (G - G_0)/A_K $.
	
\end{enumerate}
The $ G $ band extinction correction is then simply given by $(A_G/A_K) A_K$, the extinction ratio determined using the synthetic photometry multiplied by the $ A_K $ value obtained from the RJCE method. We find that our sample of red clump extinction ratios are well approximated by
\begin{align}\label{eq:AGAKfit}
    A_G/A_K = 7.09 - 2.16 A_K &+ 0.68 A_K^2 + 0.18 A_K^3 - 0.12 A_K^4 \nonumber\\
    &+ 0.49/1000\text{K} (T_\text{eff} - 4835\text{K}).
\end{align}
This parameterization is shown along with the sample and its residuals in \figurename~\ref{fig:AGAKs}.
\section{Joint Luminosity and \emph{Gaia} parallax zero point offset calibration methodology}\label{sec:Methods}

\subsection{Hierarchical modeling of the red clump in \gaia\ DR2}\label{sec:BaseModel}

We construct a probabilistic model based on previous analyses of \gaia\ observations of standard candles. In particular, our model is inspired by \citet{SesarRRLyrae}, in which RR Lyrae were used to simultaneously validate \gaia\ DR1 parallaxes and fit a luminosity function for the stars. With parallax calibration parameters $\theta_\varpi$, red clump luminosity function parameters $\theta_\text{RC}$, and distance prior parameters $ \theta_r $, the posterior probability of all the model parameters $\theta = (\theta_\varpi, \theta_\text{RC}, \theta_r)$ is then proportional to the likelihood of the data $\mathcal{D}$, and the prior probability of the model parameters.
\begin{equation}\label{eq:Posterior}
	p(\theta | \mathcal{D}) \propto p(\mathcal{D} | \theta) p(\theta).
\end{equation}
The likelihood of the entire red clump data set can be split into the parallax and magnitude likelihoods for each $i$-th star independently:
\begin{equation}\label{eq:Likelihood}
	p(\mathcal{D} | \theta) = \prod_i p(\varpi_i | \theta_\varpi) p(m_i | \theta_\text{RC})
\end{equation}

We model the parallax measurements of each red clump star in \gaia\ DR2 as being drawn from a normal distribution. In other words, the likelihood of the DR2 parallax measurements is
\begin{align}
	&p(\varpi_i| \theta_\varpi) \sim \mathcal{N} \big( \varpi_i \big| \varpi_i', \varsigma_{\varpi_i}^2 \big) \text{, where}\label{eq:PlxProb}\\
	&\mathcal{N} \big(x \big| \mu_x, \sigma_x^2 \big) = \frac{1}{\sqrt{2\pi\sigma_x^2}} e^{-\frac{(x-\mu_x)^2}{2\sigma_x^2}},\label{eq:NormalDist}\\
	&\varpi_i' = 1 / r_i + \varpi_0 \text{, and}\label{eq:TruePlx}\\
	&\varsigma_{\varpi_i}^2 = (f_\varpi \sigma_{\varpi_i})^2 + \sigma_{\varpi,\text{+}}^2.\label{eq:PlxErrorValidation}
\end{align}
The distribution for each star is centered around its ``true observed parallax" $\varpi_i'$ set by \eqnname~\eqref{eq:TruePlx}, which is the inverse of its true heliocentric distance $r_i$ summed with the systematic parallax zero point offset $\varpi_0$. This is the parallax that \gaia\ would observe for a source taking into account the parallax zero point in the limit of no other measurement uncertainties. We later consider the significance of a non-constant parallax zero point $\varpi_0 (T_\text{eff}, G, \alpha, \delta)$ that is dependent on colour, magnitude, and sky position. The uncertainty for each parallax measurement is assumed to be Gaussian, with $\sigma_{\varpi_i}^2$ being the reported parallax uncertainty from \gaia\ DR2, but we allow for adjustments to the reported uncertainties to account for mis-estimated uncertainties. \eqnname~\eqref{eq:PlxErrorValidation} includes two error correction parameters $f_\varpi$, and $\sigma_{\varpi,\text{+}}^2$, which were used to inflate the Tycho-Gaia Astrometric Solution (TGAS) parallax uncertainties reported in \gaia\ DR1 \citep{TGAS}. The values used in \gaia\ DR1 were $f_\varpi = 1.4$, and $\sigma_{\varpi,\text{+}} = 0.2\ \text{mas}$. The reported values for \gaia\ DR2 are $f_\varpi = 1.08$, $\sigma_{\varpi,+}(G < 13) = 0.021\ \text{mas}$, and $\sigma_{\varpi,+}(G > 13) = 0.043\ \text{mas}$\footnote{\url{https://www.cosmos.esa.int/documents/29201/1770596/Lindegren_GaiaDR2_Astrometry_extended.pdf}}. We include these parameters $\theta_\varpi = \{\varpi_0, f_\varpi, \sigma_{\varpi,+}\}$ in our model as a validation of the reported parallax uncertainties and corrections in \gaia\ DR2. It is also of note that the astrometric solutions for bright ($ G < 13 $) and dim ($ G \geq 13 $) sources are different \citep{GaiaDR2CatValid}; thus, we also consider a few models with separate \gaia\ systematic parameters $ \theta_\varpi = \{ \varpi_0, f_\varpi, \sigma_{\varpi,+} \} $ for bright/dim sources.

The red clump requires a model for the luminosity as a function of intrinsic stellar properties such as metallicity, and colour. The absolute magnitude $M$ of the $i$-th red clump star can be described as
\begin{equation}\label{eq:RCLuminosity}
	M_i = M_\text{ref} + \alpha \big([J_0 - K_0]_i - [J_0 - K_0]_\text{ref} \big) + \beta \big( [\text{Fe/H}]_i - [\text{Fe/H}]_\text{ref} \big),
\end{equation}
where $(J_0 - K_0)$ is the extinction corrected colour. $M_\text{ref}$ is a reference value parameter which represents the absolute magnitude of a typical red clump star. Both terms denoted with a subscript ``ref" are \textit{fixed} representative values of the red clump population. In every one of our models, we use the median of each corresponding property in the sample: $(J_0 - K_0)_\text{ref} = 0.60~\text{mag}$, and $[\text{Fe/H}]_\text{ref} = -0.12~\text{dex}$.

\begin{figure}
	\centering
	\includegraphics[width=\hsize]{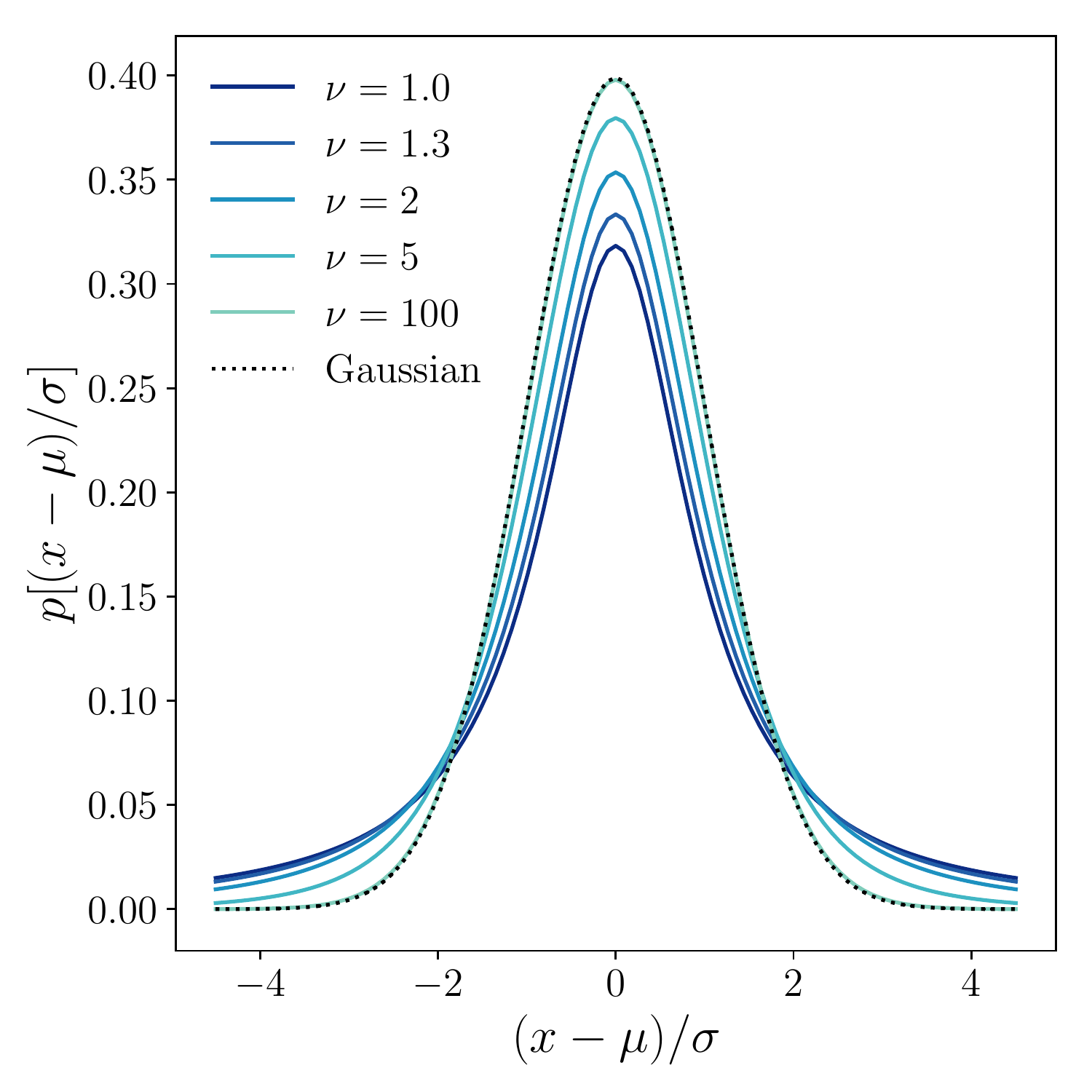}
	\caption{The Student's t-distribution in comparison with a Gaussian distribution for different values of $\nu$. The dashed line shows a standard normal distribution. For large values of $\nu$ the distribution is approximately Gaussian (the $\nu=100$ distribution essentially overlaps the Gaussian curve), but for small values of $\nu$ the distribution has much heavier tails than a Gaussian. We use the Student's t-distribution as a flexible model for the luminosity function of the red clump.}
	\label{fig:Student-t}
\end{figure}

We then choose to model the absolute magnitude of each red clump star as being drawn from a Student's $t$-distribution centered around $M_\text{ref}$, or
\begin{align}
	&p \Big( M_i \Big| [\text{Fe/H}]_i, (J_0 - K_0)_i \Big) \sim \mathcal{S} \Big( M_i \Big| M_\text{ref}, \sigma_M^2, \nu \Big) \text{, where}\label{eq:RCProbStudentt}\\
	&\mathcal{S}(t_i | \nu) = \frac{\Gamma \left( \frac{\nu+1}{2} \right)}{\sqrt{\nu\pi} \Gamma \left (\frac{\nu}{2} \right)} \left( 1 + \frac{t_i^2}{\nu} \right)^{-\frac{\nu+1}{2}} \text{, and}\label{eq:Studentt}\\
	&t_i = \frac{M_i - M_\text{ref}}{\sigma_M}.\nonumber
\end{align}
Here, $\nu$ is a parameter that controls the shape of the distribution. The role of $\nu$ in the Student's t-distribution is illustrated in \figurename~\ref{fig:Student-t}. A lower value of $\nu$ introduces a positive excess kurtosis to the distribution; in other words, the peak at the central value $M_\text{ref}$ gets narrower, and the tails become elevated. This allows the luminosity calibration to be less susceptible to outliers which still contaminate the red clump sample after quality cuts. The Student's $t$-distribution converges to a normal distribution as $\nu \rightarrow \infty$. The uncertainties associated with $m_0$, $[\text{Fe/H}]$, $(J_0 - K_0)$, and the luminosity model chosen in \eqnname~\eqref{eq:RCLuminosity} are captured in $\sigma_M^2$ as a model parameter, which then sets the width of the distribution.

To tie the luminosity and parallax calibration models together, we unite the observed magnitude of each star with its observed parallax through the distance to the star, $r_i$. This was already done for the parallaxes through \eqnname~\eqref{eq:TruePlx}, and can be done for the magnitudes through the distance modulus,
\begin{equation}
	\mu_i = m_{0,i} - M_i = 5\log_{10}r_i - 5, \label{eq:DistanceModulus}
\end{equation}
where $m_0$ is the extinction corrected apparent magnitude. When using the $ G $ magnitudes in particular, we use the corrected $ G $ magnitudes using \eqnname~\eqref{eq:Gcorr}, and apply the extinction correction from our model. We then model the likelihood of the observed distance modulus as a Student's t-distribution centered around the true distance modulus, or
\begin{equation}\label{eq:DMDist}
	p \Big( \mu_i(m_i) \Big| \mu_i(r_i), \sigma_M^2 \Big) = \mathcal{S} \Big( \mu_i(m_i) \Big| \mu_i(r_i), \sigma_M^2, \nu \Big).
\end{equation}
The approximation from \eqnname~\eqref{eq:RCProbStudentt} to \eqnname~\eqref{eq:DMDist} is valid only if measurement uncertainties for $ m_0 $, $ (J_0 - K_0) $ and $ [\text{Fe/H}] $ are sufficiently small compared to $ \sigma_M^2 $. We provide a discussion for this later on.

The true distance to each star is unknown, so we include each true distance $ r_i $ as a model parameter under an exponentially decreasing volume density prior with scale distance $L$ as described by \citet{BailerJonesPrior}:
\begin{equation}\label{eq:DistancePrior}
	p(r_i|L) = \frac{r_i^2}{2L^3}\exp(-r_i/L).
\end{equation}
The likelihood for the entire data set $\mathcal{D}$ is then
\begin{equation}\label{eq:FullLikelihood}
	p(\mathcal{D} | \theta) = \prod_i \mathcal{N}(\varpi_i | \varpi_i', \varsigma_{\varpi_i}^2) \mathcal{S}\Big( \mu_i(m_i) \Big| \mu_i(r_i), \sigma_M^2, \nu \Big)p(r_i|L).
\end{equation}
\begin{figure}
	\centering
	\includegraphics[width=\hsize]{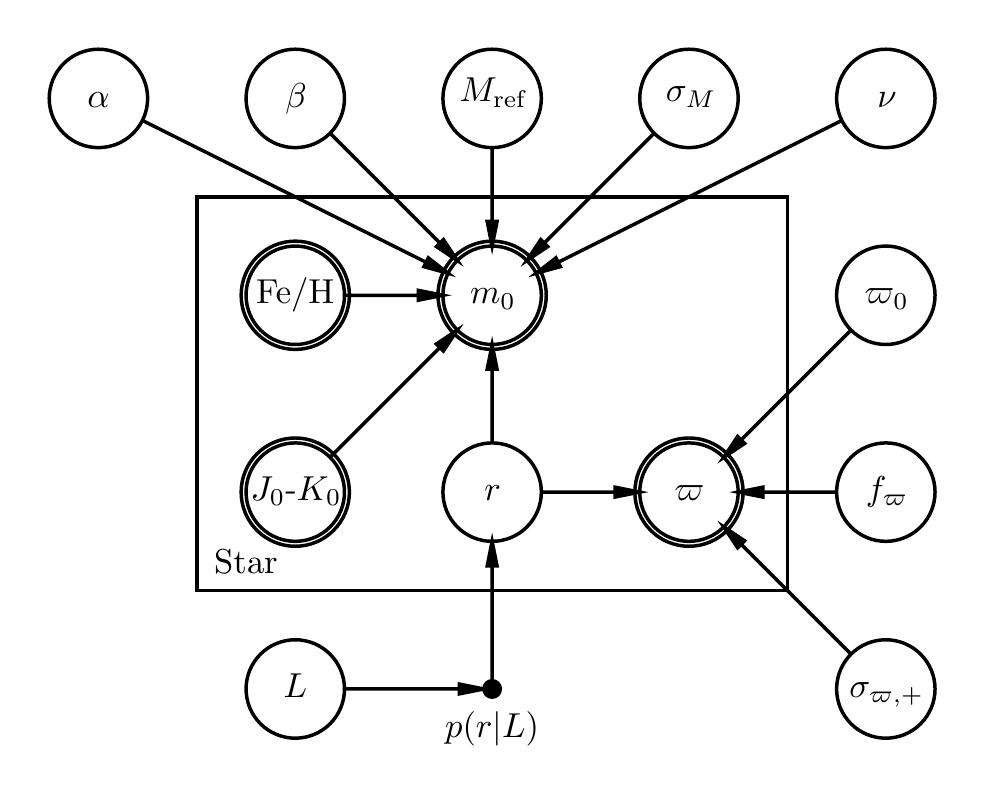}
	\caption{A probabilistic graphical model illustrating the base luminosity and parallax calibration for red clump stars. Double circled nodes indicate observed parameters (or likelihoods), and single circled nodes indicate fit parameters (which require priors). Nodes inside the rectangle are different for each star. The direction of each arrow indicates the conditional dependence of each parameter. For example, the arrow pointing from $r$ to $m_0$ indicates $p(m_0|r)$.}
	\label{fig:PGMbase}
\end{figure}
A schematic of our probabilistic model illustrating the dependencies of each parameter and observed quantity is shown in \figurename~\ref{fig:PGMbase}. Each of the single circled parameters require prior probabilities. The prior for $r$ is described by \eqnname~\eqref{eq:DistancePrior}. We assign the following broad priors for the remaining parameters:

\begin{itemize}
	\item Zero point parallax: Uniform prior between $-100 < \varpi_0/\mu\mathrm{as} < 100$
	\item \gaia\ parallax error scaling: Uniform prior between $0.2 < f_\varpi < 2$
	\item \gaia\ parallax error offset: Log-uniform prior with $0.1 < \sigma_{\varpi,+}/\mu\mathrm{as}$ and no upper bound
	\item Distance prior scale length for \eqnname~\eqref{eq:DistancePrior}: Uniform prior between $300 < L/\mathrm{pc} < 3000$
	\item Colour $(J_0 - K_0)$ slope for red clump luminosity: Uniform prior between $-5 < \alpha < 5$
	\item Metallicity $[\mathrm{Fe/H}]$ slope for red clump luminosity: Uniform prior between $-5 < \beta/(\mathrm{mag~dex}^{-1}) < 5$
	\item Reference absolute magnitude for red clump stars in the $K_s$ band: Uniform prior between $-2 < M_{K,\mathrm{ref}}/\mathrm{mag} < -1$
	\item Reference absolute magnitude for red clump stars in the $J$ band: Uniform prior between $-2.5 < M_{J,\mathrm{ref}}/\mathrm{mag} < -1.5$
	\item Reference absolute magnitude for red clump stars in the $H$ band: Uniform prior between $-2 < M_{H,\mathrm{ref}}/\mathrm{mag} < -1$
	\item Reference absolute magnitude for red clump stars in the \gaia\ $G$ band: Uniform prior between $0 < M_{G,\mathrm{ref}}/\mathrm{mag} < 1$
	\item Spread in red clump luminosities in each photometric band: Log-uniform prior between $0.01 < \sigma_{M_m}/\mathrm{mag} < 0.8$
	\item Student's $t$-distribution degrees of freedom parameter for each photometric band: Log-uniform prior between $0 < \nu < 1000$
\end{itemize}

\subsection{Multivariate Photometry Models}

The \gaia\ parallax validation model described so far uses only one photometric measurement for each star. In general, $k$ apparent magnitudes may be included in a single analysis through the multivariate $ t $-distribution
\begin{equation}\label{eq:JointPhotometryDist}
	p \Big( \vec{\mu}_i(\vec{m}_i) \Big| \vec{\mu_i}(r_i), \mathbf{\Sigma}_m \Big) = \mathcal{S} \Big( \vec{\mu}_i(\vec{m}_i) \Big| \vec{\mu_i}(r_i), \mathbf{\Sigma}_M, \nu \Big) \text{, where}
\end{equation}
\begin{equation}\label{eq:MultiStudentt}
	\mathcal{S} \Big( \vec{x} \Big| \vec{\mu}, \mathbf{\Sigma}, \nu \Big) = \frac{\Gamma[(\nu + k)/2]}{\Gamma(\nu/2)\sqrt{\nu^k\pi^k|\mathbf{\Sigma}|}} \left[ 1 + \frac{1}{\nu} (\vec{x} - \vec{\mu})^T \mathbf{\Sigma}^{-1} (\vec{x} - \vec{\mu}) \right].
\end{equation}
Each component of $ \vec{\mu}_i(\vec{m}_i) $ can be a distance modulus computed with one of the photometric bands using the left-hand side of \eqnname~\eqref{eq:DistanceModulus}, and each component of $ \vec{\mu}_i(r_i) $ is the true distance modulus computed with the distance using the right-hand side of \eqnname~\eqref{eq:DistanceModulus}. As with the single-variable Student's $t$, $ \mathbf{\Sigma} $ is related to the width of the distribution like $ \sigma_M $ while also taking into account covariances across each photometric band.

We implement the multivariate $t$-distribution to create a model which accepts both $ K_s $ and $ G $ band photometry simultaneously. This model should act as a check for the \gaia\ $ G $-band extinction model discussed in \S\ref{sec:ExtinctionModel}. For any given star, photometric measurements in different bands are expected to be highly correlated, meaning that we do not expect to infer the model parameters to higher precision with the inclusion of multiple photometric information. Rather, this multivariate model serves as a method to validate the extinction estimates for $ A_G $ described in \S\ref{sec:ExtinctionModel}, which are expected to be less accurate than the given $ A_K $ values. We specifically choose the $ K_s $ information to match with $ G $ because it is expected to be the least extinguished as the reddest photometric band available.

We also consider a different model which is more intuitively comparable to the single photometry models. While the luminosity of the red clump in each band is expected to be highly correlated, the luminosity of each star is not expected to be as strongly correlated with the intrinsic colour. We can therefore model photometry in different bands (say, $ G $ and $ K_s $) by modeling the absolute magnitude distribution of one band (say, $ K_s $) with one Student's t-distribution and the colour distribution (say, $ G - K_s $) with another Student's t-distribution, with the latter given by
\begin{align}\label{eq:ColourDist}
	p ( G_{0,i}-K_{0,i} | M_{G,i} -& M_{K,i}, \sigma_{GK}^2 ) \nonumber \\
	&= \mathcal{S} ( G_{0,i}-K_{0,i} | M_{G,i} - M_{K,i}, \sigma_{GK}^2 ),
\end{align}
where $ M_{K,i} $ and $ M_{G,i} $ are still modeled with separate versions of \eqnname~\eqref{eq:RCLuminosity}. This version of the multiple photometry model is implemented by simply including \eqnname~\eqref{eq:ColourDist} in the likelihood set by \eqnname~\eqref{eq:FullLikelihood}.

\begin{figure}
	\centering
	\includegraphics[width=\hsize]{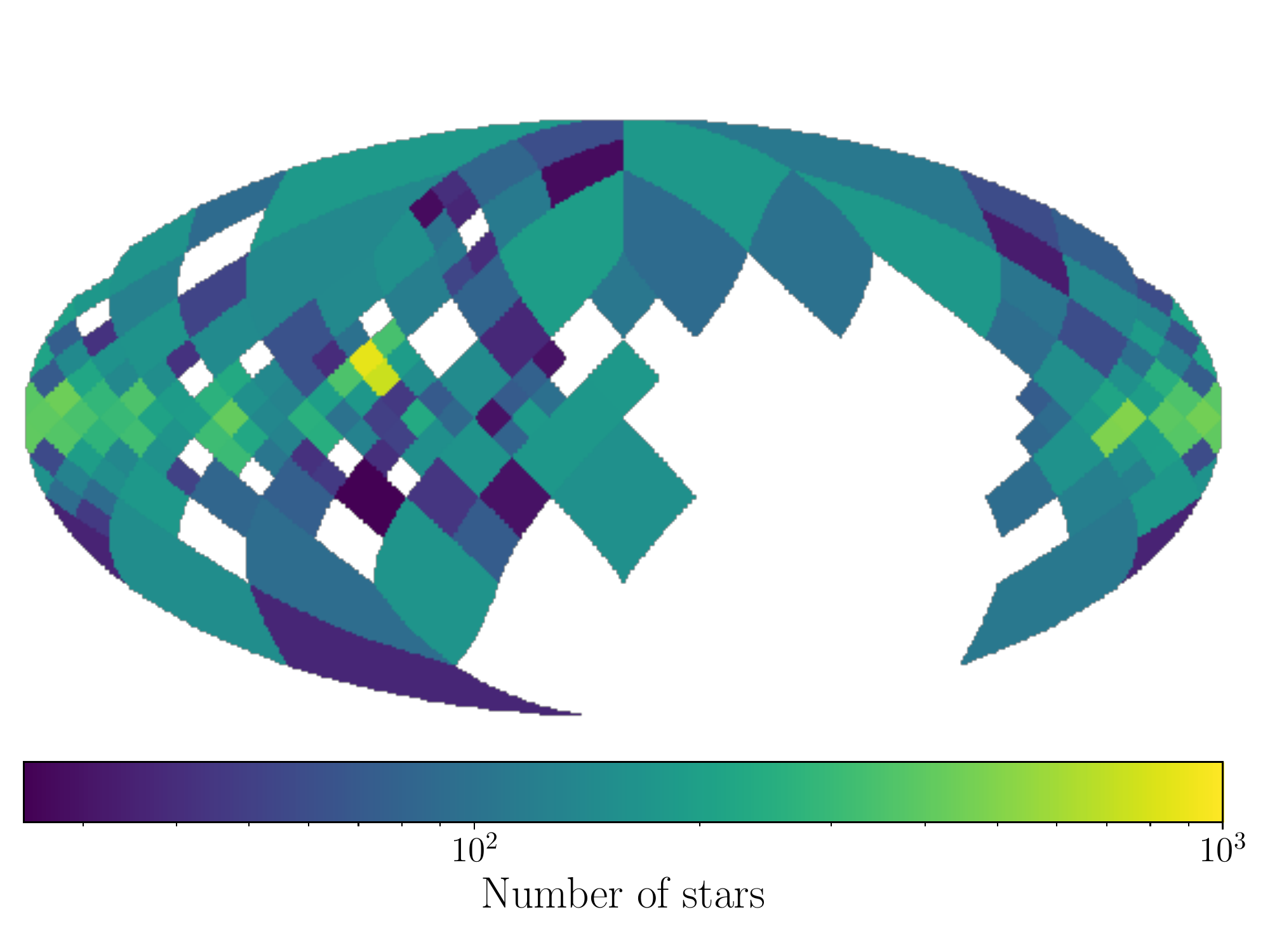}
	\caption{The \texttt{HEALPIX} projection of the sky distribution of the red clump sample divided into varying resolutions through conditions described in \S\ref{sec:VariationMethods}. We use this to determine the parallax zero point offset's variation on the sky.\label{fig:SkyMap}}
\end{figure}

\subsection{Variation of the \gaia\ Zero Point Parallax}\label{sec:VariationMethods}

We consider various ways in which the \gaia\ parallax zero point may depend on other quantities. To account for the differences in astrometric solutions for sources with $ G < 13 $ and $ G \geq 13 $, we first implement a single photometry model with separate parallax-related parameters $ \theta_\varpi = \{ \varpi_0, f_\varpi, \sigma_{\varpi,+} \} $ for each of the two cases. This and every other model discussed in this section uses only the $ K_s $ photometric information to calibrate the red clump luminosity model.

As mentioned previously, the \gaia\ parallax zero point should exhibit continuous and multivariate dependences on properties such as the observed magnitude $ G $, the observed colour $ G_{BP} - G_{RP} $, and the position of the source on the sky. Note that we investigate dependencies of the parallax zero point with respect to the \textit{uncorrected} (using neither \eqnname~\eqref{eq:Gcorr} nor the extinction corrections) $ G $ magnitudes because we want to model how the DR2 pipeline leaves systematic residuals in the astrometry. The photometric/colour dependencies are easily included in the model by introducing a functional form for $ \varpi_0 $:
\begin{equation}\label{eq:ZPVariation}
\varpi_0 = z(G, G_{BP}-G_{RP}) = z_0 +  z_G(G) + z_c(G_{BP}-G_{RP}).
\end{equation}
In particular, we investigate simple quadratic parameterizations for the dependences on both $ G $, and $ G_{BP} - G_{RP} $. Due to the change in the astrometric processing at $G = 13$, we also consider models with independent zero point functional parameterizations for sources $ G < 13 $ and $ G \geq 13 $.

We further consider models in which the \gaia\ parallax parameters $ \theta_\varpi = \{ \varpi_0, f_\varpi, \sigma_{\varpi,+} \} $ are not required to follow any specific functional parameterization, but rather we model them as separate constants in binned $ G $ or $ G_{BP}-G_{RP} $ space.  In particular, we consider 17 bins of width 0.5 mag from 9.5 -- 18 mag in $ G $ space, and we consider 7 bins of width 0.5 mag from 1 -- 4.5 mag in $ G_{BP} - G_{RP} $ space. Stars outside of these ranges are discarded for each respective analysis. For each of these binned models, we also consider separate $ L $ parameters for each bin to account for the distance dependence of $ G $, or any possible 3D position clustering of red clump stars of similar observed colour.
The dependence on sky position can be probed by projecting the red clump sample onto \texttt{HEALPIX} maps \citep{HEALPIX}. The \texttt{HEALPIX} framework divides the surface of a sphere into 12 diamond-shaped patches of equal solid angle at lowest order (called \texttt{NSIDE}, and beginning at 1). Higher resolution patches on the sphere are obtained by further dividing each patch into 4. Each division of map patches into 4 increases the \texttt{NSIDE} of the map by a factor of 2. The zero point dependence on sky position can then be modelled through unique values of $ \varpi_0 $, and $ L $ for each \texttt{HEALPIX} patch on a sky map. In particular, we consider maps of order $ \texttt{NSIDE} = \{ 1, 2, 4, 8 \} $, corresponding to patches of approximately $ \{ 3438, 860, 215, 54 \} $ square degrees respectively. The equal-area property of each patch in a \texttt{HEALPIX} projection is useful for regularly sampling the variation of $ \varpi_0 $ across the sky, but the effectiveness of this method is restricted by the need for a sufficiently large subsample of red clump stars within each patch. Much of the red clump sample is located within the Galactic disk, so the most precise measurements of $ \varpi_0 $ on the sky will come from the Galactic disk.

In order to retrieve similar quality inferences of the parallax zero point across the sky, we have developed a method for analyzing a single sky map with varying \texttt{HEALPIX} resolutions dependent on the number of stars within each patch. \texttt{HEALPIX} patches are recursively split into higher resolution sub-patches if the parent patch contains greater than 200 stars. Each daughter sub-patch is further broken up if it still contains greater than 200 stars and is of order \texttt{NSIDE} $ < 8 $. If the sub-patch contains between 25 -- 200 stars or has reached \texttt{NSIDE} $ = 8 $, then it is kept at that resolution. If the sub-patch contains fewer than 25 stars, then it is discarded. The resulting star density map across our sample can be seen in Galactic coordinates in \figurename~\ref{fig:SkyMap}. The highest resolution patches allow for detailed analysis of the zero point's variation along the Galactic plane, while the lower resolution patches should group enough stars away from the Galactic plane together to provide reasonable inferences of the parallax zero point. All parameters remain global with the exception of $ \varpi_0 $ and $ L $, which are specific to each patch. Again, we also consider a model with independent sky variations of $ \varpi_0 $ for sources $ G < 13 $ and $ G \geq 13 $.

\subsection{Red Clump Luminosity Calibration: Dependence on [$\mathbf{\alpha}$/Fe]}\label{sec:aFeMethod}

\begin{figure}
	\centering
	\includegraphics[width=\hsize]{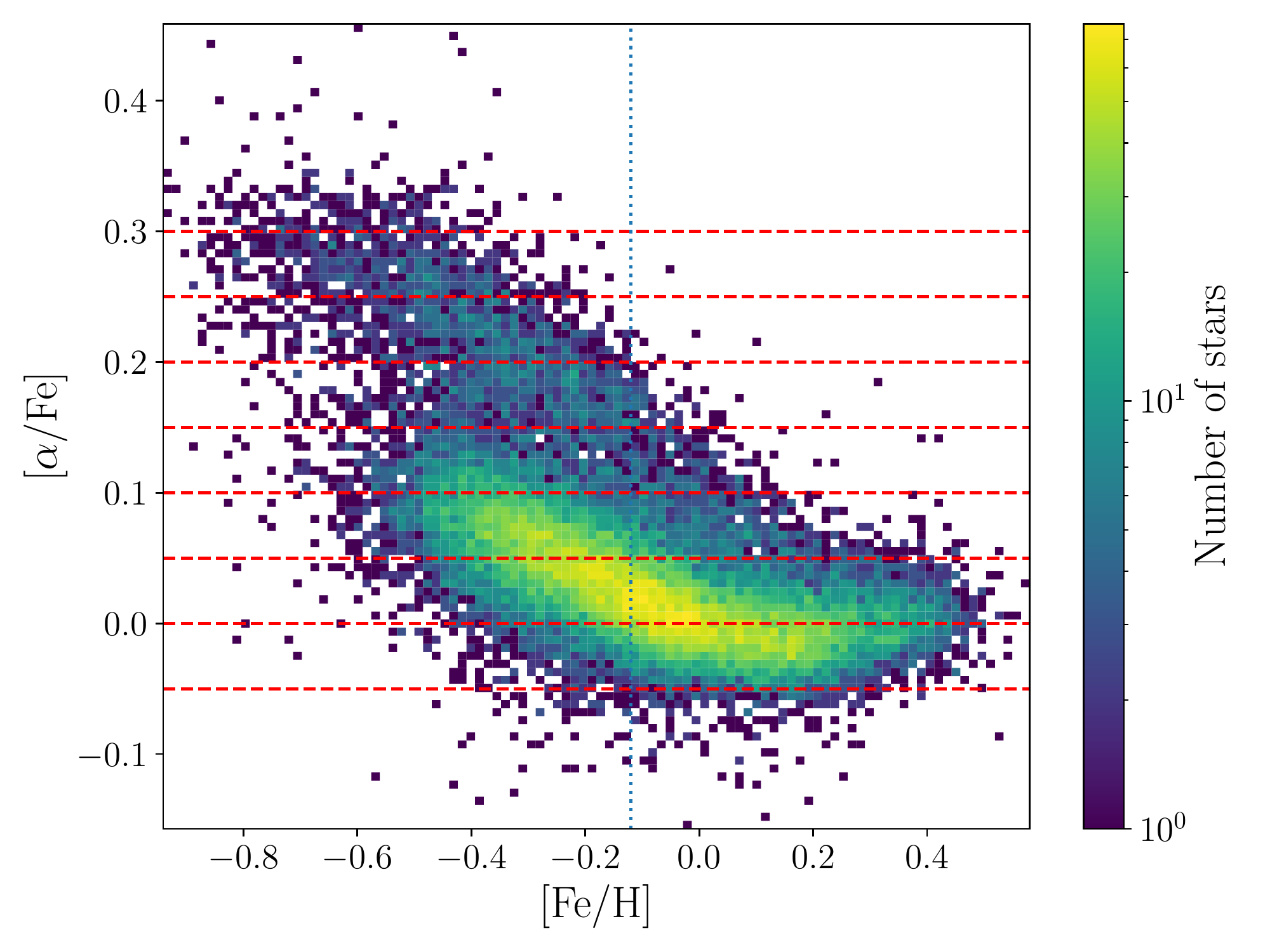}
	\caption{The distribution of red clump sample stars in chemical abundance space. The dotted blue line indicates the median metallicity [Fe/H] of the full sample. The boundaries of the $ [\alpha/\text{Fe}] $ bins described in \S\ref{sec:aFeMethod} are shown as dashed red lines. These bins are used in the analysis of variations of the red clump luminosity calibration across sub-populations.\label{fig:aFeDist}}
\end{figure}

To determine the red-clump luminosity function for sub-populations within the red clump sample, we separate the data into bins in $ [\alpha/\text{Fe}] $ (calculated with \eqnname~\eqref{eq:alphaFe} using APOGEE abundance data) space. The sample is split into 7 evenly sized bins from $[\alpha/\text{Fe}] = -0.05~\text{dex}$ to $ 0.3~\text{dex} $ that are $ 0.05~\text{dex} $ wide. The chemical distribution of the entire sample can be seen in \figurename~\ref{fig:aFeDist} along with the boundaries of the $ [\alpha/\text{Fe}] $ bins described. The low-alpha ($[\alpha/\text{Fe}] \lesssim 0.15\,\text{dex}$) and high-alpha ($[\alpha/\text{Fe}] \gtrsim 0.15\,\text{dex}$) sequences can clearly be seen in abundance space.

To test possible differences in the luminosity calibration between red clump stars of varying chemical compositions, we alter the model described by \figurename~\ref{fig:PGMbase} to include independent red clump luminosity parameters $ \theta_\text{RC} = \{ \alpha, \beta, M_\text{ref}, \sigma_M, \nu \} $ for each $ [\alpha/\text{Fe}] $ bin while the remaining parameters are still relevant to the global sample. Keep in mind that we are still taking into account the red clump luminosity dependence on $ (J_0 - K_0) $ and $ [\text{Fe/H}] $ separately for each $ [\alpha/\text{Fe}] $ bin through \eqnname~\eqref{eq:RCLuminosity}. We expect small changes in the red clump luminosity calibration for each sub-population, which can be seen as variations of $ \theta_\text{RC} $ parameters in $ [\alpha/\text{Fe}] $ space.

\subsection{Tests of red clump stellar evolution models}\label{sec:APOGEEVerifyMethod}

\citet{BovyRC} determined distances to the red clump in our sample using colour and metallicity trends determined from PARSEC stellar models \citep{PARSEC}, applying a constant calibration offset obtained from a Hipparcos red clump sample. We test the stellar model used in \citet{BovyRC} as follows: We adjust \eqnname~\eqref{eq:DistanceModulus} to
\begin{equation}\label{eq:StellarModelValidationDM}
	\mu_i = m_{0,i} - M_i', 
\end{equation}
where
\begin{equation}\label{eq:StellarModelValidation}
	M_i' = M_i\big( [J_0-K_0]_i, [\text{Fe/H}]_i \big) + m_{0,i} - \mu_i^*,
\end{equation}
and $M_i^* = m_{0,i} - \mu_i^*$ is the absolute magnitude of a red clump star predicted with the stellar model. The new observed distance modulus which will replace \eqnname~\eqref{eq:DistanceModulus} for this validation model is then
\begin{equation}\label{eq:DMObservedStellarModel}
	\mu_i = \mu_i^* - M_i.
\end{equation}
Here, $ M_i $ still represents a red clump luminosity calibrated with the probabilistic model, but it is not tied to any photometric band. Instead, it is related to the accuracy of the stellar models used to determine $ \mu_i^* $ in \citet{BovyRC}. We expect $M_i \rightarrow 0$, with each parameter within \eqnname~\eqref{eq:RCLuminosity} now describing possible residuals of the stellar model; i.e, the parameters $ M_\text{ref} $, $ \alpha $, and $ \beta $ are expected to be 0 if the stellar models used in the catalogue describe the red clump at least as well as the empirical models in this paper. 

\subsection{Implementation of Models}
\begin{figure}
	\centering
	\includegraphics[width=\hsize]{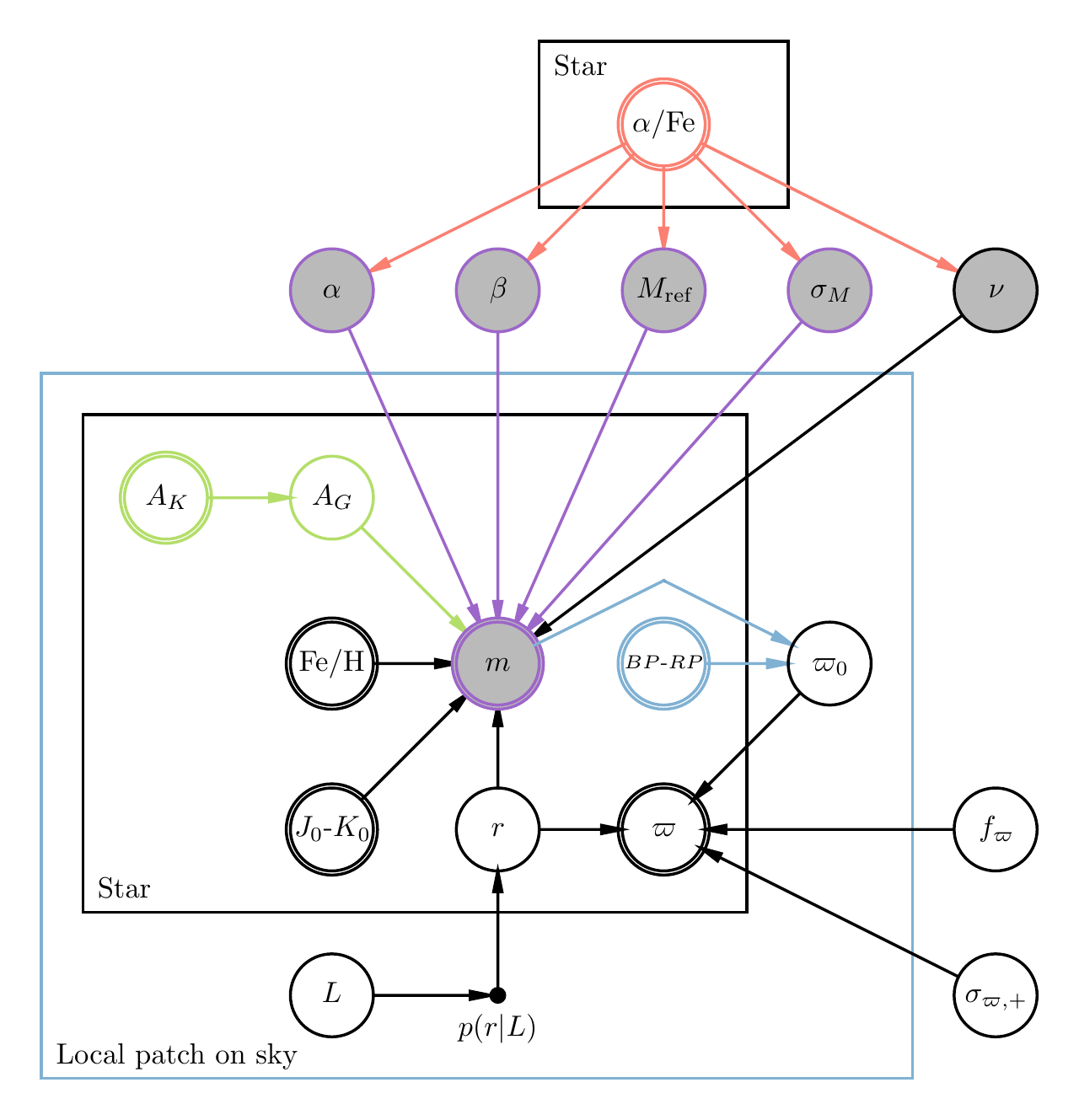}
	\caption{A probabilistic graphical model illustrating the extensions to the base luminosity and parallax calibration of red clump stars \figurename~\ref{fig:PGMbase}. The colours illustrate an independent add-on to the existing base model shown in \figurename~\ref{fig:PGMbase}. \textit{Blue}: Adding functional dependencies to the parallax zero point $\varpi_0$, or modeling the spatial variations of $ \varpi_0 $ across the sky with \texttt{HEALPIX} patches. \textit{Green}: Adding an extinction model for the \gaia\ $G$ band photometry. \textit{Red}: Adding a $[\alpha/\mathrm{Fe}]$ dependence of the red clump luminosity calibration set by \eqnname~\eqref{eq:RCLuminosity}. \textit{Purple}: Changing the red clump luminosity calibration to a validation of the distance moduli reported in the red clump catalogue \citep{BovyRC}. \textit{Gray fill}: Including multiple photometric bands simultaneously in the analysis. A unique copy of each of these parameters is added for each corresponding photometric band.}
	\label{fig:PGMfull}
\end{figure}
A full probabilistic graphical model in \figurename~\ref{fig:PGMfull} illustrates each extension to the base model in \figurename~\ref{fig:PGMbase}. The sheer number of parameters that need to be sampled for each model (especially the distances to $ \sim 28000 $ stars) hinders the use of traditional Markov chain Monte Carlo (MCMC) techniques. We choose instead to use \texttt{Stan}, a Hamiltonian Monte Carlo (HMC) software \citep{Stan}. Some advantages of using HMC instead of MCMC include a generally more efficient sampling of parameter space, especially in the case of highly correlated parameters. HMC also provides more accurate and precise sampling of parameter space for both models with complex distributions and models with many parameters. The principles and benefits of HMC are further described in great detail by \citet{HMC}.

Each of the models described in this paper was implemented in the \texttt{Python 3} wrapper for \texttt{Stan}, also known as \texttt{PyStan}\footnote{Stan Development Team. 2018. PyStan: the Python interface to Stan, Version 2.17.1.0. \url{http://mc-stan.org}}. The posterior distribution space for each model was sampled with 4 chains for 2500 steps, and the first 1000 steps of each chain were discarded as warm-up steps. In total, 6000 samples were collected for the posterior distributions of each model we report. The parameters were intialized randomly according to their respective priors with the exception of the distances. The true distance parameter for each star was initialized using the absolute inverse of its parallax reported in \gaia\ DR2.

\afterpage{
\clearpage
\begin{landscape}
	\begin{table}
		\caption{Inferred model parameters for all models discussed in this paper along with $ \pm 1\sigma $ uncertainties. Adjacent rows from the same model/analysis are indicated with shared background highlighting.}
		\label{tab:results}
		\renewcommand{\arraystretch}{1.25}
		\begin{tabular}{lcccccccccc}
			\hline
			Photometric band & $ \varpi_0 $ & $ f_\varpi $ & $ \sigma_{\varpi,+} $ & $ M_\text{ref} $ & $ \sigma_M $ & $ \alpha $ & $ \beta $ & $ \nu $ & FWHM$/2.355$ & $ L $ \\
			and model & [$\mu$as] & -- & [$\mu$as] & [mag] & [mag] & -- & [mag/dex] & --  & [mag] & [pc] \\
			\hline
			
			$ K_s $ (Base) & $ -47.87^{+0.79}_{-0.76}  $ & $ 1.46^{+0.01}_{-0.01} $ & $ 0.44 ^{+0.90}_{-0.28} $ & $ -1.622^{+0.004}_{-0.004} $ & $ 0.097^{+0.003}_{-0.003} $ & $ 0.24^{+0.07}_{-0.07} $ & $ -0.21^{+0.01}_{-0.01} $ & $ 1.29^{+0.03}_{-0.03} $ & $ 0.088^{+0.003}_{-0.003} $ & $ 987^{+4}_{-4} $ \\
			\rowcolor{Gray}
			$ G $  (Base) & $ -38.30^{+0.88}_{-0.80} $ & $ 1.49^{+0.01}_{-0.01} $ & $ 0.54^{+1.41}_{-0.37} $ & $ 0.447^{+0.004}_{-0.005} $ & $ 0.123^{+0.003}_{-0.003} $ & $ 2.92^{+0.08}_{-0.08} $ & $ -0.11^{+0.02}_{-0.02} $ & $ 1.48^{+0.03}_{-0.03} $ & $ 0.116^{+0.004}_{-0.004} $ & $ 1030^{+4}_{-4} $ \\
			
			$ J $ (Base) & $ -47.91^{+0.79}_{-0.76} $ & $ 1.46^{+0.01}_{-0.01} $ & $ 0.41^{+0.87}_{-0.25} $ & $ -1.019^{+0.004}_{-0.004} $ & $ 0.097^{+0.003}_{-0.003} $ & $ 1.25^{+0.08}_{-0.07} $ & $ -0.21^{+0.01}_{-0.01} $ & $ 1.29^{+0.03}_{-0.03} $ & $ 0.088^{+0.003}_{-0.003} $ & $ 987^{+4}_{-4} $ \\
			\rowcolor{Gray}
			$ H $ (Base) & $ -49.69^{+0.82}_{-0.81} $ & $ 1.46^{+0.01}_{-0.01} $ & $ 0.41^{+0.85}_{-0.25} $ & $ -1.516^{+0.004}_{-0.004} $ & $ 0.098^{+0.003}_{-0.003} $ & $ 0.50^{+0.07}_{-0.08} $ & $ -0.19^{+0.01}_{-0.01} $ & $ 1.30^{+0.03}_{-0.03} $ & $ 0.089^{+0.004}_{-0.004} $ & $ 981^{+4}_{-4} $ \\
			
			$ K_s $ (Base; Fixed $ f_\varpi $, $ \sigma_{\varpi,+} $) & $ -50.25^{+0.86}_{-0.85} $ & $ 1.08 $ & $ {}^{21, G < 13}_{43, G > 13} $ & $ -1.613^{+0.004}_{-0.004} $ & $ 0.093^{+0.002}_{-0.002} $ & $ 0.28^{+0.08}_{-0.07} $ & $ -0.22^{+0.01}_{-0.01} $ & $ 1.27^{+0.03}_{-0.02} $ & $ 0.084^{+0.003}_{-0.003} $ & $ 983^{+4}_{-4} $\\
			\rowcolor{Gray}
			$ K_s $ (Joint $ K_s $ \& $ G $; Multi-$t$) & $ -48.94^{+0.93}_{-0.96} $ & $ 1.45^{+0.01}_{-0.01} $ & $ 0.41^{+0.72}_{-0.25} $ & $ -1.628^{+0.005}_{-0.005} $ & $ {}^a 0.041^{+0.001}_{-0.001} $ & $ -0.02^{+0.09}_{-0.09} $ & $ -0.19^{+0.02}_{-0.02} $ & $ 2.74^{+0.05}_{-0.05} $ & $ {}^a 0.063^{+0.001}_{-0.001} $ & $ 984^{+4}_{-4} $ \\
			\rowcolor{Gray}
			$ G $ (Joint $ K_s $ \& $ G $; Multi-$t$) & $ -48.94^{+0.93}_{-0.96} $ & $ 1.45^{+0.01}_{-0.01} $ & $ 0.41^{+0.72}_{-0.25} $ & $ 0.508^{+0.005}_{-0.005} $ & $ {}^a 0.049^{+0.001}_{-0.001} $ & $ 2.73^{+0.09}_{-0.09} $ & $ -0.11^{+0.02}_{-0.02} $ & $ 2.74^{+0.05}_{-0.05} $ & $ {}^a 0.069^{+0.001}_{-0.001} $ & $ 984^{+4}_{-4} $ \\
			
			$ K_s $ (Joint $ K_s $ \& $ G $; $ G - K_s $) & $ -47.92^{+0.84}_{-0.81} $ & $ 1.46^{+0.01}_{-0.01} $ & $ 0.43^{+0.90}_{-0.28} $ & $ -1.622^{+0.004}_{-0.004} $ & $ 0.097^{+0.003}_{-0.003} $ & $ 0.25^{+0.07}_{-0.07} $ & $ -0.21^{+0.01}_{-0.01} $ & $ 1.29^{+0.03}_{-0.03} $ & $ 0.088^{+0.003}_{-0.003} $ & $ 987^{+4}_{-4} $ \\
			
			$ G $  (Joint $ K_s $ \& $ G $; $ G - K_s $) & $ -47.92^{+0.84}_{-0.81} $ & $ 1.46^{+0.01}_{-0.01} $ & $ 0.43^{+0.90}_{-0.28} $ & $ 0.517^{+0.004}_{-0.004} $ & $ 0.159^{+0.001}_{-0.001} $ & $ 3.03^{+0.08}_{-0.08} $ & $ -0.12^{+0.01}_{-0.01} $ & $ 2.58^{+0.03}_{-0.03} $ & $ 0.165^{+0.002}_{-0.002} $ & $ 987^{+4}_{-4} $ \\
			\rowcolor{Gray}
			$ K_s $ (Joint $ K_s $ \& $ J $; $ J - K_s $) & $ -47.89^{+0.82}_{-0.83} $ & $ 1.46^{+0.01}_{-0.01} $ & $ 0.42^{+0.89}_{-0.27} $ & $ -1.622^{+0.004}_{-0.004} $ & $ 0.097^{+0.003}_{-0.003} $ & $ 0.25^{+0.07}_{-0.08} $ & $ -0.21^{+0.01}_{-0.01} $ & $ 1.29^{+0.03}_{-0.03} $ & $ 0.088^{+0.003}_{-0.003} $ & $ 987^{+4}_{-4} $ \\
			\rowcolor{Gray}
			$ J $  (Joint $ K_s $ \& $ J $; $ J - K_s $) & $ -47.89^{+0.82}_{-0.83} $ & $ 1.46^{+0.01}_{-0.01} $ & $ 0.42^{+0.89}_{-0.27} $ & $ -1.020^{+0.004}_{-0.004} $ & $ 0.098^{+0.003}_{-0.003} $ & $ 1.25^{+0.07}_{-0.08} $ & $ -0.21^{+0.01}_{-0.01} $ & $ 1.51^{+0.03}_{-0.03} $ & $ 0.093^{+0.001}_{-0.001} $ & $ 987^{+4}_{-4} $ \\
			
			$ K_s $ ($ G < 13 $) & $ -35.74^{+1.55}_{-1.53} $ & $ 1.23^{+0.03}_{-0.03} $ & $ 0.51^{+1.16}_{-0.34} $ & $ -1.665^{+0.006}_{-0.006} $ & $ 0.103^{+0.003}_{-0.003} $ & $ 0.34^{+0.07}_{-0.08} $ & $ -0.20^{+0.01}_{-0.01} $ & $ 1.34^{+0.03}_{-0.03} $ & $ 0.095^{+0.004}_{-0.004} $ & $ 581^{+4}_{-4} $ \\
			
			$ K_s $ ($ G \geq 13 $) & $ -42.36^{+0.84}_{-0.86} $ & $ 1.52^{+0.01}_{-0.01} $ & $ 0.49^{+1.12}_{-0.32} $ & $ -1.665^{+0.006}_{-0.006} $ & $ 0.103^{+0.003}_{-0.003} $ & $ 0.34^{+0.07}_{-0.08} $ & $ -0.20^{+0.01}_{-0.01} $ & $ 1.34^{+0.03}_{-0.03} $ & $ 0.095^{+0.004}_{-0.004} $ & $ 1262^{+7}_{-6} $ \\
			\rowcolor{Gray}
			$ K_s $ ($ G < 13 $; Fixed $ f_\varpi $, $ \sigma_{\varpi,+} $) & $ -35.39^{+1.57}_{-1.54} $ & $ 1.08 $ & $ 21 $ & $ -1.666^{+0.006}_{-0.005} $ & $ 0.094^{+0.003}_{-0.003} $ & $ 0.32^{+0.08}_{-0.08} $ & $ -0.20^{+0.01}_{-0.01} $ & $ 1.28^{+0.03}_{-0.03} $ & $ 0.086^{+0.003}_{-0.003} $ & $ 581^{+4}_{-4} $ \\
			\rowcolor{Gray}
			$ K_s $ ($ G \geq 13 $; Fixed $ f_\varpi $, $ \sigma_{\varpi,+} $) & $ -44.75^{+0.87}_{-0.86} $ & $ 1.08 $ & $ 43 $ & $ -1.666^{+0.006}_{-0.005} $ & $ 0.094^{+0.003}_{-0.003} $ & $ 0.32^{+0.08}_{-0.08} $ & $ -0.20^{+0.01}_{-0.01} $ & $ 1.28^{+0.03}_{-0.03} $ & $ 0.086^{+0.003}_{-0.003} $ & $ 1261^{+6}_{-6} $ \\
			
			$ K_s $ ($ G < 13 $; Fixed $ M_\text{ref} $) & $ -46.68^{+0.63}_{-0.63} $ & $ 1.20^{+0.03}_{-0.02} $ & $ 0.52^{+1.13}_{-0.35} $ & $ -1.622 $ & $ 0.104^{+0.003}_{-0.003} $ & $ 0.33^{+0.07}_{-0.07} $ & $ -0.21^{+0.01}_{-0.01} $ & $ 1.34^{+0.03}_{-0.03} $ & $ 0.096^{+0.004}_{-0.004} $ & $ 569^{+3}_{-3} $ \\
			
			$ K_s $ ($ G > 13 $; Fixed $ M_\text{ref} $) & $ -48.17^{+0.46}_{-0.44} $ & $ 1.51^{+0.01}_{-0.01} $ & $ 0.46^{+1.02}_{-0.30} $ & $ -1.622 $ & $ 0.104^{+0.003}_{-0.003} $ & $ 0.33^{+0.07}_{-0.07} $ & $ -0.21^{+0.01}_{-0.01} $ & $ 1.34^{+0.03}_{-0.03} $ & $ 0.096^{+0.004}_{-0.004} $ & $ 1236^{+5}_{-5} $ \\
			\rowcolor{Gray}
			$ K_s $ ($ G $ dep.; All $ G $; Fixed $ M_\text{ref} $) & $ {}^b -60.91 $ & $ 1.42^{+0.01}_{-0.01} $ & $ 0.41^{+0.80}_{-0.25} $ & $ -1.622 $ & $ 0.104^{+0.003}_{-0.003} $ & $ 0.15^{+0.07}_{-0.07} $ & $ -0.21^{+0.01}_{-0.01} $ & $ 1.31^{+0.03}_{-0.02} $ & $ 0.095^{+0.004}_{-0.004} $ & $ 986^{+3}_{-3} $ \\
			
			$ K_s $ ($ G $ dep.; $ G < 13 $; Free $ M_\text{ref} $) & $ {}^b 72.51 $ & $ 1.31^{+0.03}_{-0.02} $ & $ 0.52^{+1.21}_{-0.35} $ & $ -1.943^{+0.013}_{-0.013} $ & $ 0.111^{+0.003}_{-0.003} $ & $ 0.04^{+0.08}_{-0.08} $ & $ -0.12^{+0.02}_{-0.02} $ & $ 1.38^{+0.03}_{-0.03} $ & $ 0.103^{+0.004}_{-0.004} $ & $ 660^{+5}_{-6} $ \\
			
			$ K_s $ ($ G $ dep.; $ G \geq 13 $; Free $ M_\text{ref} $) & $ {}^b -24.31 $ & $ 1.53^{+0.01}_{-0.01} $ & $ 0.56^{+1.54}_{-0.38} $ & $ -1.943^{+0.013}_{-0.013} $ & $ 0.111^{+0.003}_{-0.003} $ & $ 0.04^{+0.08}_{-0.08} $ & $ -0.12^{+0.02}_{-0.02} $ & $ 1.38^{+0.03}_{-0.03} $ & $ 0.103^{+0.004}_{-0.004} $ & $ 1433^{+11}_{-11} $ \\
			\rowcolor{Gray}
			$ K_s $ ($ G $ dep.; $ G < 13 $; Fixed $ M_\text{ref} $) & $ {}^b -42.93 $ & $ 1.20^{+0.03}_{-0.03} $ & $ 0.50^{+1.14}_{-0.33} $ & $ -1.622 $ & $ 0.106^{+0.003}_{-0.003} $ & $ 0.30^{+0.07}_{-0.07} $ & $ -0.21^{+0.01}_{-0.01} $ & $ 1.34^{+0.03}_{-0.03} $ & $ 0.098^{+0.004}_{-0.004} $ & $ 569^{+3}_{-3} $ \\
			\rowcolor{Gray}
			$ K_s $ ($ G $ dep.; $ G \geq 13 $; Fixed $ M_\text{ref} $) & $ {}^b -60.90 $ & $ 1.49^{+0.01}_{-0.01} $ & $ 0.46^{+1.05}_{-0.29} $ & $ -1.622 $ & $ 0.106^{+0.003}_{-0.003} $ & $ 0.30^{+0.07}_{-0.07} $ & $ -0.21^{+0.01}_{-0.01} $ & $ 1.34^{+0.03}_{-0.03} $ & $ 0.098^{+0.004}_{-0.004} $ & $ 1236^{+5}_{-6} $ \\
			
			$ K_s $ ($ G $ dep.; $ G < 13 $; $ M_\text{ref} $ prior) & $ {}^b -29.20 $ & $ 1.20^{+0.02}_{-0.03} $ & $ 0.52^{+1.17}_{-0.34} $ & $ -1.658^{+0.003}_{-0.003} $ & $ 0.106^{+0.003}_{-0.003} $ & $ 0.30^{+0.08}_{-0.08} $ & $ -0.21^{+0.01}_{-0.01} $ & $ 1.34^{+0.03}_{-0.03} $ & $ 0.098^{+0.004}_{-0.004} $ & $ 579^{+4}_{-4} $ \\
			
			$ K_s $ ($ G $ dep.; $ G \geq 13 $; $ M_\text{ref} $ prior) & $ {}^b -56.58 $ & $ 1.50^{+0.01}_{-0.01} $ & $ 0.48^{+1.04}_{-0.32} $ & $ -1.658^{+0.003}_{-0.003} $ & $ 0.106^{+0.003}_{-0.003} $ & $ 0.30^{+0.08}_{-0.08} $ & $ -0.21^{+0.01}_{-0.01} $ & $ 1.34^{+0.03}_{-0.03} $ & $ 0.098^{+0.004}_{-0.004} $ & $ 1256^{+6}_{-6} $ \\
			\rowcolor{Gray}
			$ K_s $ ($ G $ dep.; Binned; Fixed $ M_\text{ref} $) & $ {}^c -47.60^{+0.37}_{-0.37} $ & $ {}^c 1.39^{+0.01}_{-0.01} $ & $ {}^c 1.56^{+0.59}_{-0.59} $ & $ -1.622 $ & $ 0.108^{+0.003}_{-0.003} $ & $ 0.31^{+0.07}_{-0.08} $ & $ -0.21^{+0.01}_{-0.01} $ & $ 1.35^{+0.03}_{-0.03} $ & $ 0.100^{+0.004}_{-0.004} $ & $ {}^c 603^{+2}_{-2} $ \\
			
			$ K_s $ ($ BP-RP $ dep.; All $ G $) & $  {}^b -100.33 $ & $ 1.43^{+0.01}_{-0.01} $ & $ 0.44^{+0.97}_{-0.28} $ & $ -1.651^{+0.005}_{-0.004} $ & $ 0.101^{+0.003}_{-0.003} $ & $ 0.13^{+0.07}_{-0.07} $ & $ -0.13^{+0.01}_{-0.01} $ & $ 1.34^{+0.03}_{-0.03} $ & $ 0.093^{+0.004}_{-0.004} $ & $ 998^{+4}_{-4} $ \\
			\rowcolor{Gray}
			$ K_s $ ($ BP-RP $ dep.; $ G < 13 $) & $ {}^b -204.21 $ & $ 1.26^{+0.03}_{-0.03} $ & $ 0.48^{+1.11}_{-0.32} $ & $ -1.658^{+0.006}_{-0.006} $ & $ 0.105^{+0.003}_{-0.003} $ & $ 0.16^{+0.07}_{-0.08} $ & $ -0.08^{+0.02}_{-0.02} $ & $ 1.41^{+0.03}_{-0.03} $ & $ 0.098^{+0.004}_{-0.004} $ & $ 574^{+4}_{-4} $ \\
			\rowcolor{Gray}
			$ K_s $ ($ BP-RP $ dep.; $ G \geq 13 $) & $ {}^b -106.98 $ & $ 1.50^{+0.01}_{-0.01} $ & $ 0.51^{+1.16}_{-0.34} $ & $ -1.657^{+0.006}_{-0.006} $ & $ 0.105^{+0.003}_{-0.003} $ & $ 0.16^{+0.07}_{-0.08} $ & $ -0.08^{+0.02}_{-0.02} $ & $ 1.41^{+0.03}_{-0.03} $ & $ 0.098^{+0.004}_{-0.004} $ & $ 1256^{+7}_{-6} $ \\
			
			$ K_s $ ($ BP-RP $ dep.; Binned) & $ {}^c -49.40^{+0.58}_{-0.58} $ & $ {}^c 1.31^{+0.01}_{-0.01} $ & $ {}^c 1.52^{+0.99}_{-0.99} $ & $ -1.634^{+0.004}_{-0.004} $ & $ 0.106^{+0.003}_{-0.003} $ & $ 0.26^{+0.07}_{-0.07} $ & $ -0.19^{+0.01}_{-0.01} $ & $ 1.37^{+0.03}_{-0.03} $ & $ 0.098^{+0.004}_{-0.004} $ & $ {}^c 953^{+4}_{-4} $ \\
			\rowcolor{Gray}
			$ K_s $ (Sky dep.; All $ G $) & $ {}^c -43.06^{+0.37}_{-0.37} $ & $ 1.44^{+0.01}_{-0.01} $ & $ 0.00^{+0.00}_{-0.00} $ & $ -1.650^{+0.004}_{-0.005} $ & $ 0.095^{+0.003}_{-0.003} $ & $ -0.08^{+0.07}_{-0.08} $ & $ -0.14^{+0.01}_{-0.01} $ & $ 1.29^{+0.03}_{-0.03} $ & $ 0.087^{+0.003}_{-0.003} $ & $ {}^c 891^{+3}_{-3} $ \\
			
			$ K_s $ (Sky dep.; $ G < 13 $) & $ {}^c -34.92^{+0.69}_{-0.69} $ & $ 1.23^{+0.03}_{-0.03} $ & $ 0.00^{+0.00}_{-0.00} $ & $ -1.668^{+0.006}_{-0.005} $ & $ 0.101^{+0.003}_{-0.003} $ & $ -0.17^{+0.07}_{-0.08} $ & $ -0.09^{+0.01}_{-0.01} $ & $ 1.37^{+0.03}_{-0.03} $ & $ 0.093^{+0.004}_{-0.004} $ & $ {}^c 571^{+3}_{-3} $ \\
			
			$ K_s $ (Sky dep.; $ G \geq 13 $) & $ {}^c -42.04^{+0.46}_{-0.46} $ & $ 1.51^{+0.01}_{-0.01} $ & $ 0.00^{+0.00}_{-0.00} $ & $ -1.668^{+0.006}_{-0.005} $ & $ 0.101^{+0.003}_{-0.003} $ & $ -0.17^{+0.07}_{-0.08} $ & $ -0.09^{+0.01}_{-0.01} $ & $ 1.37^{+0.03}_{-0.03} $ & $ 0.093^{+0.004}_{-0.004} $ & $ {}^c 1229^{+6}_{-6} $ \\
			\rowcolor{Gray}
			$ K_s $ ($ [\alpha/\text{Fe}] $ bins) & $ -48.60^{+0.77}_{-0.80} $ & $ 1.46^{+0.01}_{-0.01} $ & $ 0.49^{+1.04}_{-0.32} $ & $ {}^c -1.629^{+0.003}_{-0.003} $ & $ {}^c 0.086^{+0.003}_{-0.003} $ & $ {}^c -0.18^{+0.08}_{-0.08} $ & $ {}^c 0.01^{+0.02}_{-0.02} $ & $ {}^c 1.29^{+0.03}_{-0.03} $ & $ 0.078^{+0.003}_{-0.003} $ & $ 985^{+4}_{-4} $ \\
			
			$\mu$ (Stellar model validation) & $ -47.21^{+0.80}_{-0.85} $ & $ 1.47^{+0.01}_{-0.01} $ & $ 0.44^{+0.89}_{-0.28} $ & $ {}^d -0.036^{+0.004}_{-0.004} $ & $ {}^d 0.105^{+0.003}_{-0.003} $ & $ {}^d 1.48^{+0.07}_{-0.07} $ & $ {}^d -0.39^{+0.01}_{-0.01} $ & $ {}^d 1.33^{+0.03}_{-0.03} $ & $ 0.096^{+0.004}_{-0.004} $ & $ 990^{+4}_{-4} $ \\
			\hline
		\end{tabular}\\
		\footnotesize 
		$ {}^a $ Diagonal components of $ \mathbf{\Sigma} $ are reported here. $ G $ and $ K_s $ share a covariance of $ \sigma_{GK} = 0.042^{+0.001}_{-0.001} $. FWHM are computed in one dimension along the zero point of the opposite band.\\
		$ {}^b $ Reported values are integrated means of the modeled/inferred functional forms of each dependence, which are further described in \S\ref{sec:VariationResults}.\\
		$ {}^c $ Reported values are computed inverse variance weighted means of each all inferred binned distributions. See \figurename~\ref{fig:zpDepBinsG} and \figurename~\ref{fig:ZPVariationsBPRP} for the binned distributions.\\
		$ {}^d $ Parameters reported here are not the same red clump luminosity calibration parameters, and a detailed interpretation is discussed in \S\ref{sec:RCModelingResults}.
	\end{table}
\end{landscape}
\clearpage
}

\section{Results}\label{sec:Results}

\subsection{Basic results}

The posterior distributions for all models are summarized in \tablename~\ref{tab:results}. We report the median and $ \pm 1\sigma $ ranges for each model parameter. It is worthwhile to note that we also obtain posterior distributions for the distances to each individual star in our sample. Individual distances are typically constrained to $ \sim10\% $, which may seem insignificant, but the combined data set allows for the parallax zero point to be inferred to $ \sim 1\% $ across every model.

\begin{figure*}
	\centering
	\includegraphics[width=\hsize]{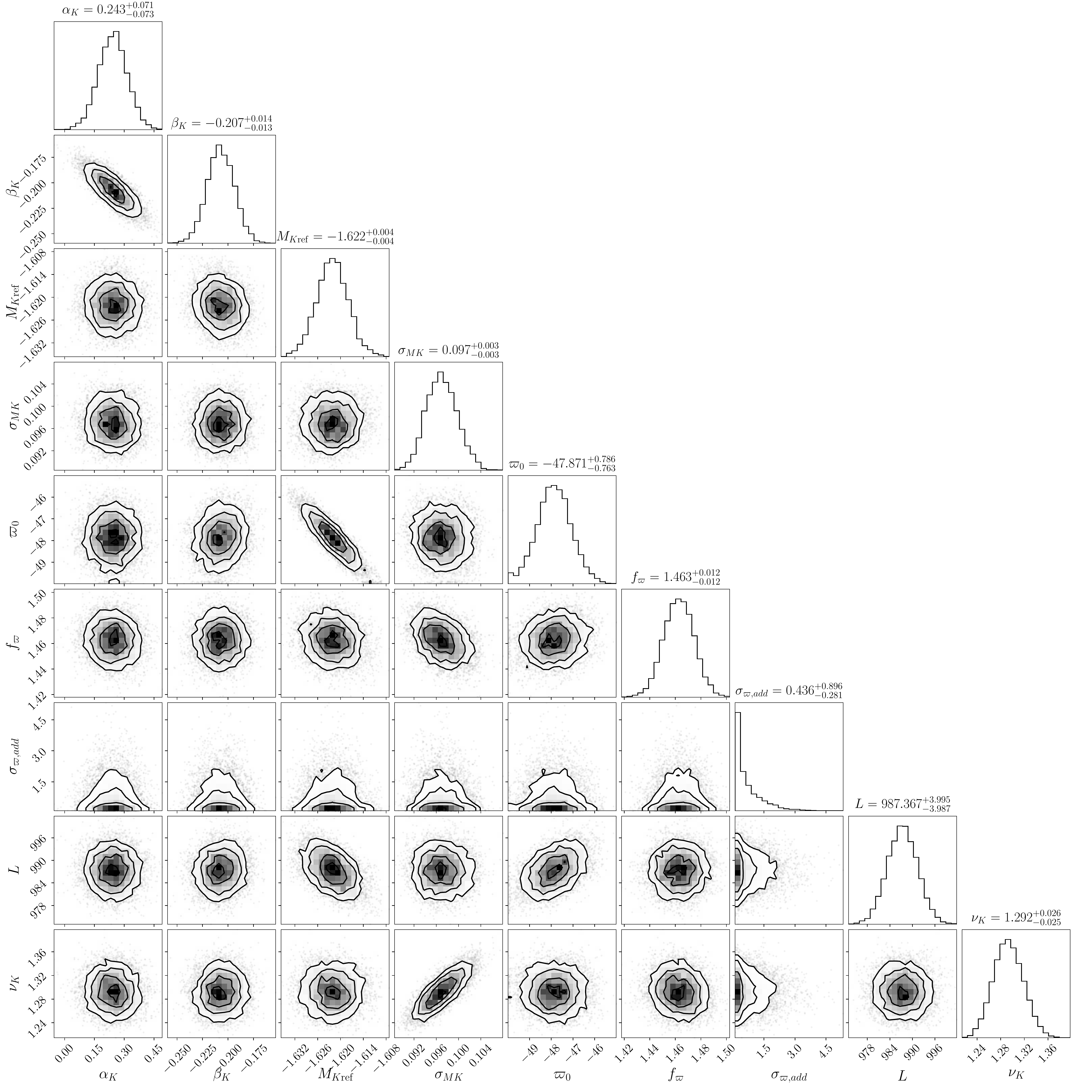}
	\caption{The posterior distributions of global parameters with respect to the $K_s$ data for constant $\varpi_0$. All parameters are precisely constrained by the data, and $ \varpi_0 $ is strongly correlated with the absolute magnitude $ M_{K,\text{ref}} $ and is weakly correlated with the distance prior parameter $ L $. The parallax zero point has close to vanishing correlations with the remaining parameters.}
	\label{fig:InitialCorner}
\end{figure*}

We find that zero points inferred across each base model using $ K_s $, $ J $, and $ H $ photometry are consistent ($ \varpi_0 = -48 \pm 1~\mu\text{as} $), with an inconsistent result from the zero point inferred with the base model using $ G $ photometry ($ \varpi_0 = -38.30^{+0.88}_{-0.80}~\mu\text{as} $). This inconsistency is alleviated with the joint photometry model using the multivariate $t$-distribution to include $ K_s $ and $ G $ data simultaneously. The resulting zero point inferred with joint photometry is $ \varpi_0 = -48.94^{+0.93}_{-0.96}~\mu\text{as} $. We also infer consistent \gaia\ parallax uncertainty correction parameters ($ f_\varpi $ and $ \sigma_{\varpi,+} $) across all base model and the joint photometry analyses. A representative posterior distribution is shown in \figurename~\ref{fig:InitialCorner} along with the correlations between each model parameter in the base model using $ K_s $ photometry.

We infer the absolute magnitude of the red clump to be $ M_\text{ref} = -1.622 \pm 0.004 $ in $ K_s $, $ M_\text{ref} = 0.447 \pm 0.005 $ in $ G $, $ M_\text{ref} = -1.019 \pm 0.004 $ in $ J $, and $ M_\text{ref} = -1.516 \pm 0.004 $ in $ H $. Red clump luminosity calibrations infer significant colour ($ J_0 - K_0 $) and metallicity ([Fe/H]) dependences in every photometric band considered. These dependences are implied again with the validation of reported APOGEE distances discussed later.

The Student's $t$-distribution appears to be able to capture outliers in the sample by widening the distribution about $ \mu_i(m_i) $; the degrees-of-freedom parameter is inferred to be approximately $ \nu \approx 1.3 $ across all models. Consequently, the scaling parameter $ \sigma_M $ does not fully describe the dispersion of stars about $ M_\text{ref} $. The Student's $t$-distribution has a formal variance of $ \infty $ for $ 1 < \nu < 2 $, but we wish to describe the width of the Student's $t$-distribution with a combination of $ \sigma_M $ and $ \nu $. We are only interested in the dispersion of stars that have not been considered outliers (which exist in the heightened tails of the distribution), so we report dispersions about the red clump luminosity calibration given by the full width at half maximum (FHWM) of each inferred distribution. \tablename~\ref{tab:results} includes the FWHM$ / 2\sqrt{2\ln(2)} $ for each red clump luminosity distribution, which is equivalent to $ 1\sigma $ for a Gaussian distribution.

Note that the spread in the red clump luminosity distribution describes the spread in distance moduli as in \eqnname~\eqref{eq:DMDist}. This includes uncertainties from photometric ($ \sigma_m  \lesssim 0.025$ mag) and metallicity ($ \sigma_\text{[Fe/H]} \lesssim 0.01 $ dex) measurements; however, these uncertainties are much smaller than the spread in the distributions (FWHM$ /2.355 \sim 0.1 $ mag). Taking this into account, the dispersion in the red clump distributions should be dominated by a combination of intrinsic dispersion and modeling uncertainties.

The inconsistencies in inferred \gaia\ parallax zero points between $ G $ and 2MASS photometry appears to be resolved with the implementation of the multiple photometry models. The inferred absolute magnitude of the red clump in $ G $ changes from $ M_G = 0.447 \pm 0.004 $ to $ M_G = 0.508 \pm 0.005 $ in the multivariate $t$ model, and to $ M_G = 0.517 \pm 0.004 $ in the model with separate single-variable $t$-distributions for $ K_s $ and $ G - K_s $. As a result, the inferred \gaia\ parallax zero point becomes consistent with the previous estimates of $ \varpi_0 \approx -49 \mu\text{as} $. This is likely an indication that our model for the $ G $ band extinction is not accurate enough for an independent analysis of the \gaia\ parallax zero point using only $ G $ photometry. Only by supplementing the model with $ K_s $ photometry are we able to achieve consistent results.

In the case of considering 2 constant zero points for $ G < 13 $ and $ G \geq 13 $, we infer $ \varpi_0 = -35.74 \pm 1.55~\mu\text{as} $ for sources $ G < 13 $ and $ \varpi_0 = -42.36 \pm 0.86~\mu\text{as} $ for sources $ G \geq 13 $. We also find once again that the \gaia\ parallax uncertainty parameters are discrepant from those reported, although we see a larger $ f_\varpi $ inferred for dim sources compared to bright sources. This is similar to the reported values of $ \sigma_{\varpi,+} $.  In an identical analysis with the exception of fixing $ f_\varpi $ and $ \sigma_{\varpi,+} $ to the reported values, we find very similar posterior distributions, implying that our probabilistic model prefers inflating \gaia\ parallax errors with $ f_\varpi $ rather than $ \sigma_{\varpi,+} $ to similar effect. We support this by running the base model analysis using $ K_s $ photometry, but adding a gamma distribution prior on $ \sigma_{\varpi,+} $:
\begin{equation}
p(\sigma_{\varpi,+} | u, v ) = \frac{v^u {\sigma_{\varpi,+}}^{u-1} e^{-v\sigma_{\varpi,+}}}{\Gamma(u)},
\end{equation}
where $ u > 0 $ and $1/(10~\mu\text{as}) < v < 1/(0.1~\mu\text{as}) $ are hyperparameters with the described priors. In particular, we find $ v $, the 'rate parameter', to hug the upper bound of the prior $ 1/(0.2^{+0.5}_{0.1}~\mu\text{as}) $. This in turn causes $ \sigma_{\varpi,+} $ to again be consistent with 0 ($ \sigma_{\varpi,+} = 1.2^{+1.2}_{-0.8}~\mu\text{as} $). All remaining parameters in this model remain consistent with the original model.

Finally, we repeat the above analysis with free $ f_\varpi $ and $ \sigma_{\varpi,+} $, but we fix $ M_\text{ref} = -1.622 $. Justification for this analysis is presented in the following section.  We infer more consistent values of $ \varpi_0 = -46.68 \pm 0.63~\mu\text{as} $ for $ G < 13 $ and $ \varpi_0 = -48.17 \pm 0.46~\mu\text{as} $ for $ G \geq 13 $.

\subsection{Variation of the \gaia\ Zero Point}\label{sec:VariationResults}

\begin{figure}
	\centering
	\includegraphics[width=\hsize]{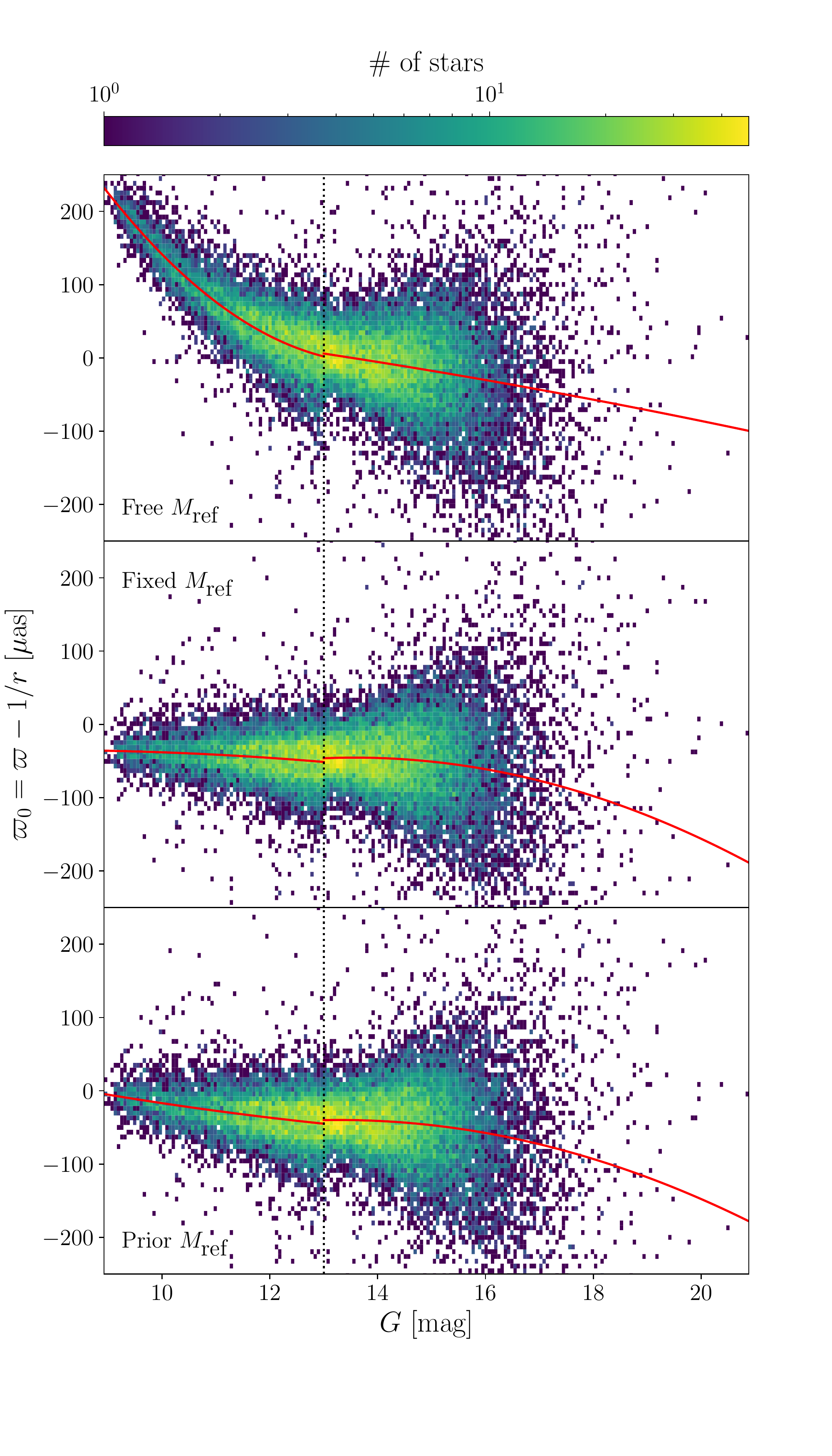}
	\vspace{-30pt}
	\caption{Inferred parallax zero point variations  $ G $. Individual estimates of the zero point for each star are shown by solving \eqnname~\eqref{eq:TruePlx} for $ \varpi_0 $, using the mean of the posterior samples of the distances to each star as well as its measured parallax. Red lines show the resulting quadratic parameterization of the zero point variations in $ G < 13 $ and $ G \geq 13 $. The model with a liberal $ M_{K,\text{ref}} $ prior appears to infer less inaccurate distances and parallax zero point offset than the other two models with more conservative priors on $ M_{K,\text{ref}} $. See \figurename~\ref{fig:RCOffset} for a comparison using distances from the APOGEE red clump catalogue \citep{BovyRC}.\label{fig:ZPVariationsG}}
\end{figure}

\begin{figure*}
\centering
\includegraphics[width=0.8\hsize]{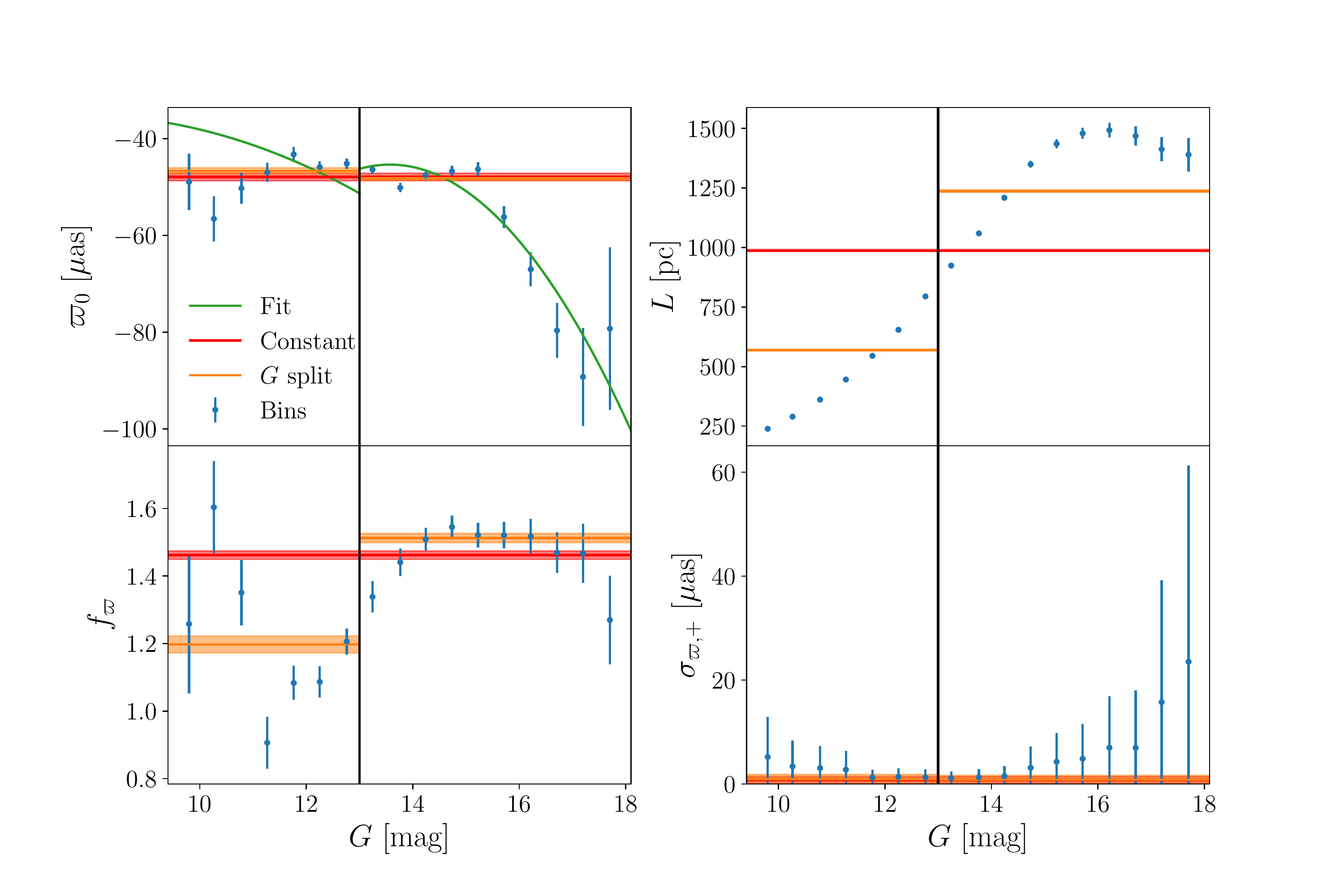}
\caption{The variations in parallax model parameters with $ G $. All results shown are from $ K_s $-based models in which $ M_\text{ref} = -1.622 $ is fixed. \textit{Blue:} Inferred model parameters across bins in $ G $. \textit{Green:} Inferred quadratic dependences for $ \varpi_0 $ as a function of $ G $. The separate parameterizations for $ G < 13 $ and $ G \geq 13 $ are shown.\textit{Red:} Inferred distributions of constant parameters for all $ G $. \textit{Orange:} Inferred parameters from a model in which parameters are considered constant in $ G $, with these four parameters being considered separately for $ G < 13 $ and $ G \geq 13 $. The parallax zero point offset is approximately constant down to $G \lesssim 15$, where the majority of our sample lives.\label{fig:zpDepBinsG}}
\end{figure*}

We begin by reporting the results from inferring a quadratic form of the zero point dependence on $ G $ separately for sources $ G < 13 $ and $ G \geq 13 $ to account for differences in astrometric solution. This was done using the $ K_s $ photometry to calibrate the red clump luminosity. We find
\begin{align}
	\varpi_0 / \mu\text{as} = &9.29^{+0.88}_{-0.91} (G - 12)^2 - 37.45^{+1.59}_{-1.57} (G - 12)\nonumber \\
	& + 30.47^{+2.92}_{-3.00} \qquad (G < 13),
\end{align}
\begin{align}
	\varpi_0 / \mu\text{as} = &-0.27^{+0.51}_{-0.49} (G - 14.5)^2 - 12.04^{+0.61}_{-0.59} (G - 14.5)\nonumber \\
	& - 11.28^{+1.46}_{-1.49} \qquad (G \geq 13).
\end{align}
This is illustrated in \figurename~\ref{fig:ZPVariationsG} in the panel labeled ``Free $ M_\text{ref} $". The inferred $ G < 13 $ fit appears to behave very differently from the constant parallax zero points inferred in the prevoius models, allowing for relatively large positive values of the zero point for sources of low $ G $. This appears to be an effect of using photometric information to infer the red clump luminosity calibration, the distances to each star, as well as the zero point dependence on $ G $ all at the same time. The fact that this model infers $ M_\text{ref} = -1.943 \pm 0.013 $ in $ K_s $ (See \tablename~\ref{tab:results}) in contrast to the previous model further suggests that this method of modeling the parallax zero point offset variation with $ G $ may be inaccurate to some degree.

To alleviate this possible degeneracy of photometric information, we implement an identical model with the exception of applying a strict prior by fixing $ M_\text{ref} = -1.622 $. We find
\begin{align}\label{eq:Gdepl13}
	\varpi_0 / \mu\text{as} = &-0.59^{+0.83}_{-0.80} (G - 12)^2 - 4.96^{+1.01}_{-1.03} (G - 12)\nonumber \\
	& -45.66^{+0.88}_{-0.86} \qquad (G < 13),
\end{align}
\begin{align}\label{eq:Gdepg13}
	\varpi_0 / \mu\text{as} = & -2.68^{+0.49}_{-0.49} (G - 14.5)^2 - 4.90^{+0.55}_{-0.54} (G - 14.5)\nonumber \\
	& - 47.62^{+0.59}_{-0.61} \qquad (G \geq 13),
\end{align}
which appears to agree with previous estimates much better. This is illustrated in \figurename~\ref{fig:ZPVariationsG} in the panel labeled ``Fixed $ M_\text{ref} $".

To demonstrate a model in which the prior on $ M_\text{ref} $ is slightly more relaxed, we repeat the above analyses. This time, $ M_\text{ref} $ is given a prior which reflects the inferred value from the base model. In other words, we implement:
\begin{equation}
	p(M_\text{ref}) = \mathcal{N}(-1.622, (0.004)^2),
\end{equation}
and we find
\begin{align}
	\varpi_0 / \mu\text{as} = & 0.55^{+0.81}_{-0.84} (G - 12)^2 - 8.82^{+1.09}_{-1.08} (G - 12)\nonumber \\
	& -36.54^{+1.24}_{-1.26} \qquad (G < 13),
\end{align}
\begin{align}
	\varpi_0 / \mu\text{as} = & -2.38^{+0.50}_{-0.49} (G - 14.5)^2 - 5.77^{+0.54}_{-0.55} (G - 14.5)\nonumber \\
	& -43.32^{+0.74}_{-0.76} \qquad (G \geq 13).
\end{align}
We infer $ M_\text{ref} = -1.658 \pm 0.003 $ from this model, and the resulting fit can be seen in \figurename~\ref{fig:ZPVariationsG} in the panel labeled ``Prior $ M_\text{ref} $". The inferred parameterization appears similar to the Fixed $ M_\text{ref} $ model, with a slightly stronger slope in $ G < 13 $. Both models appear to favour a linear solution to the $ G $ dependence of the zero point, with quadratic coefficients consistent with 0.

We also report results from a model considering a single quadratic parameterization of the zero point dependence on $ G $ for all $ G $. For this model, we fix $ M_\text{ref} = -1.622 $. We find
\begin{align}
	\varpi_0 / \mu\text{as} = &-0.52^{+0.17}_{-0.16} (G - 16)^2 - 7.65^{+0.89}_{-0.87} (G - 16)\nonumber \\
	&- 62.50^{+1.18}_{-1.10}.
\end{align}

Finally, we investigate a model in which we do not enforce a specific functional parameterization of the parallax zero point's dependence on $ G $. The results are shown in \figurename~\ref{fig:zpDepBinsG}, in which the parallax parameters $ \theta_\varpi = \{ \varpi_0, f_\varpi, \sigma_{\varpi,+} \} $ and $ L $ are modeled as separate constants along bins in $ G $ space. We fix $ M_\text{ref} = -1.622 $ in this model. The variations of $ \varpi_0 $ with $ G $ appear to be well modeled by the quadratic parametrization for $ G \geq 13 $, while it appears to be better parameterized as a constant for $ G < 13 $.

\begin{figure}
	\centering
	\includegraphics[width=\hsize]{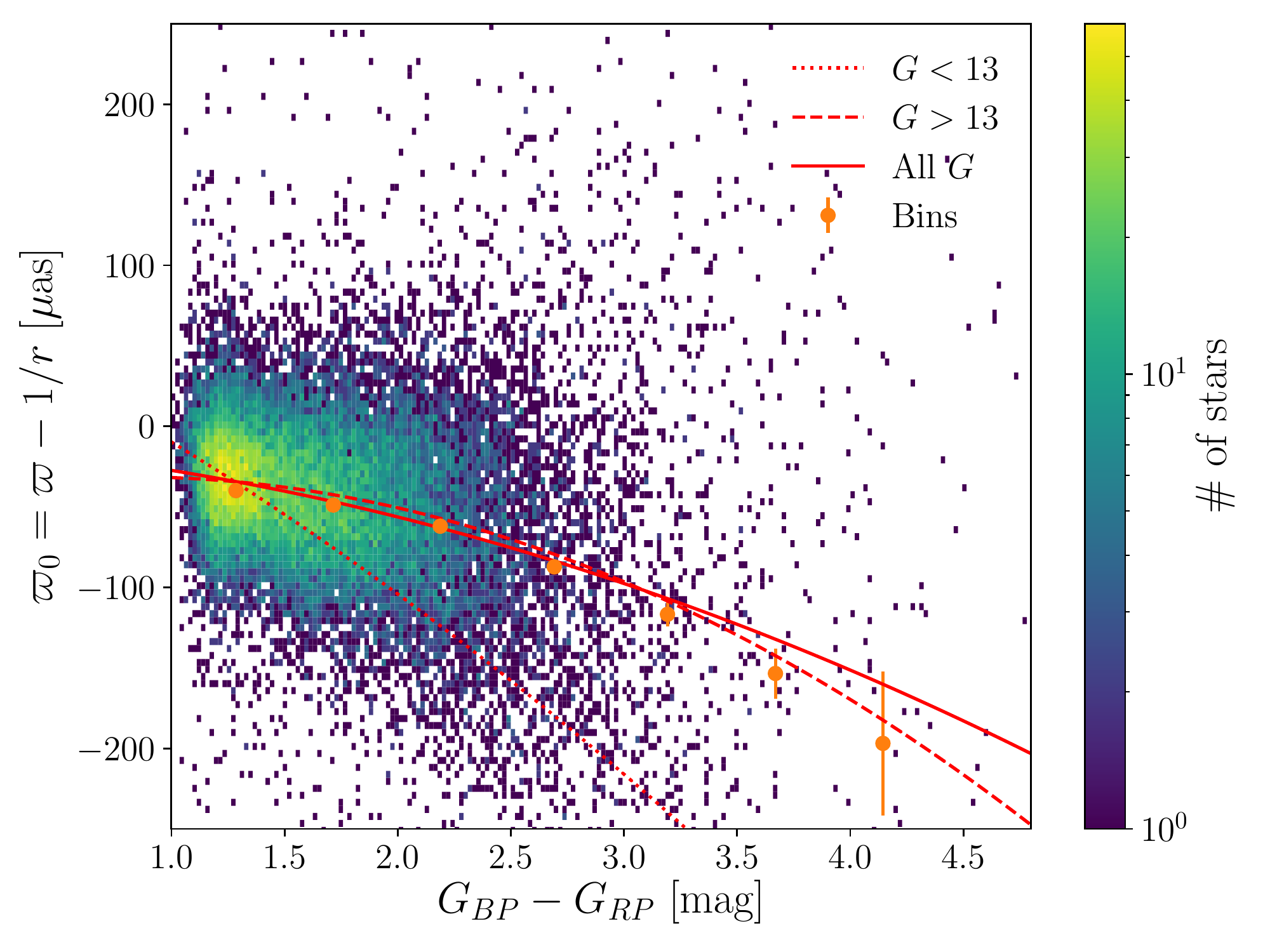}
	\caption{Variations of the \gaia\ parallax zero point with observed colour. \textit{Red:} The quadratic parameterizations of the variation. \textit{Orange:} The parallax zero point across colour bins for all $ G $. A density of individual estimates from each star using \eqnname~\eqref{eq:TruePlx} with measured parallax and distance posteriors from the ``All $ G $" quadratic parameterization model is shown in the background. The $ G < 13 $ quadratic parameterization appears to not be a good model of the parallax zero point variation with colour; whereas, the remaining models agree with one another.\label{fig:ZPVariationsBPRP}}
\end{figure}

\begin{figure}
	\centering\vspace{-24pt}
	\includegraphics[width=\hsize]{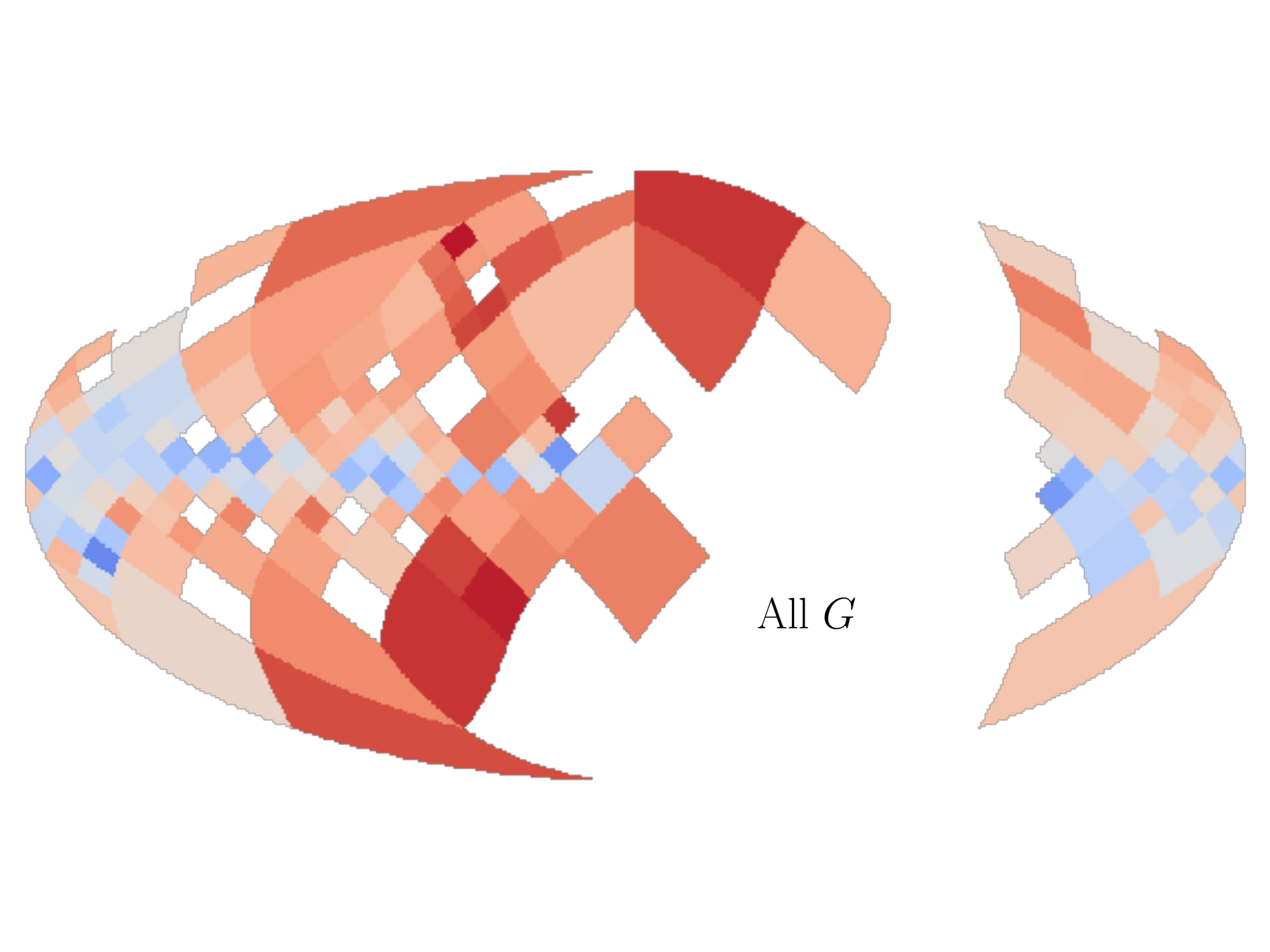}\\ \vspace{-24pt}
	\includegraphics[width=\hsize]{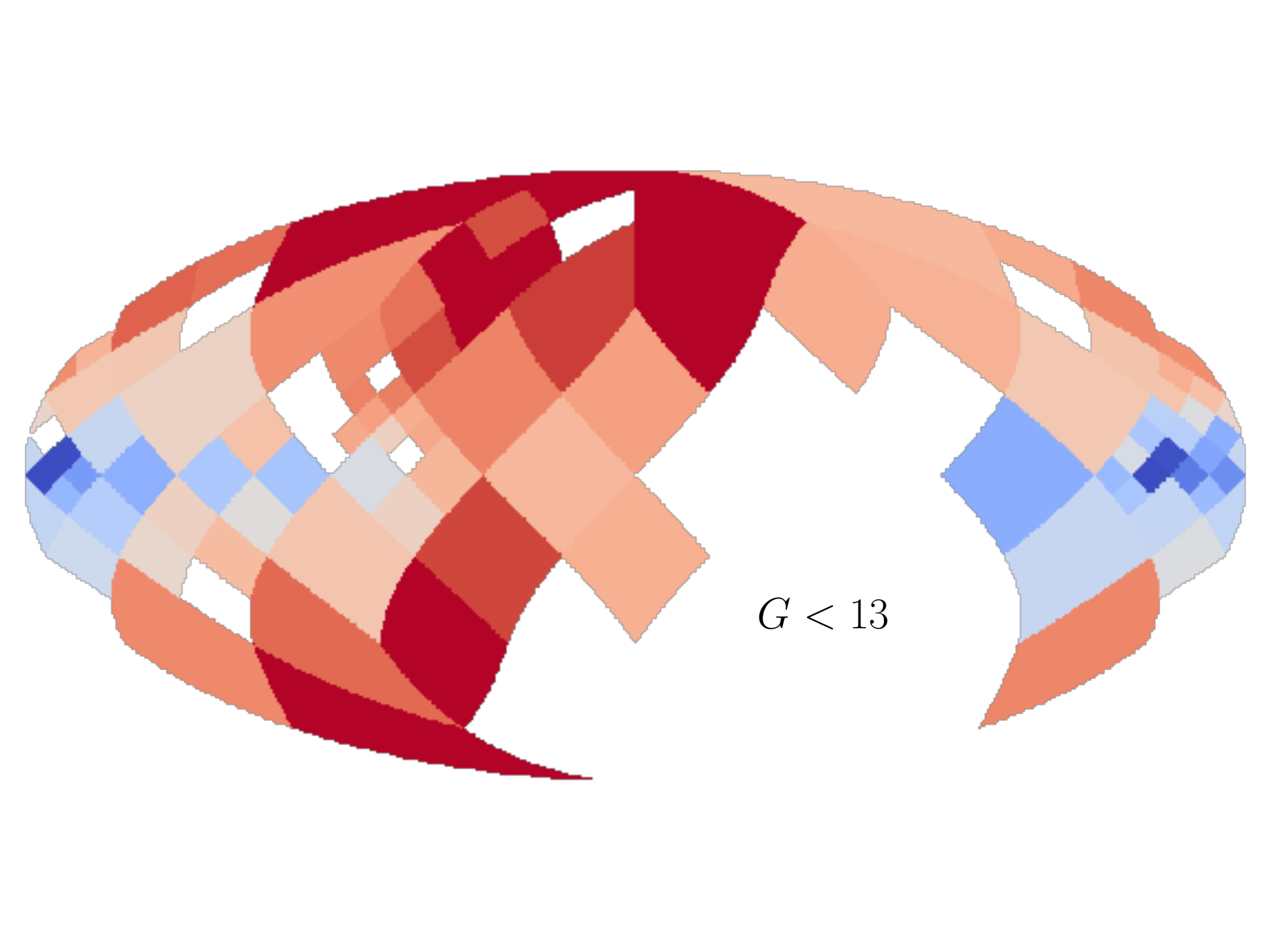}\\ \vspace{-24pt}
	\includegraphics[width=\hsize]{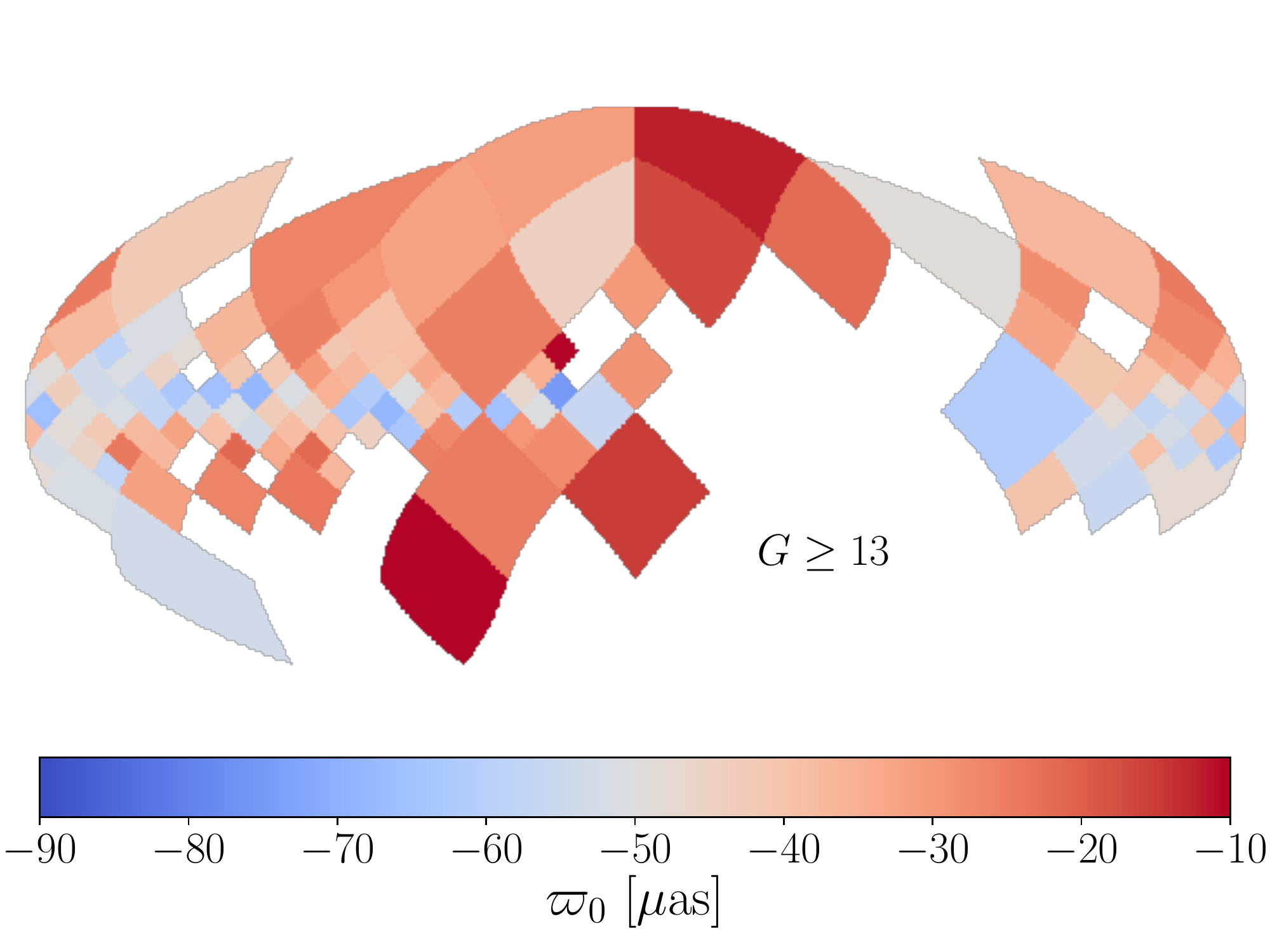}
	\caption{Variations of the \gaia\ parallax zero point across the sky, shown here in Galactic coordinates. Patches with $ \varpi_0 / \sigma_{\varpi_0} < 3 $ have been masked. \textit{Top:} Inferred zero points for all $ G $. \textit{Middle:} Inferred zero points for $ G < 13 $. \textit{Bottom:} Inferred zero points for $ G \geq 13 $.\label{fig:ZeroPointMap}}
\end{figure}

The \gaia\ parallax zero point also appears to exhibit variations with respect to observed colour. The following models allow $ M_\text{ref} $ to be free, as the observed colour and magnitude of the sources should not share the same degeneracies as in the $ G $ dependent models. In the quadratic parameterization for all sources, we find
\begin{align}\label{eq:BPRPdepAllG}
	\varpi_0 / \mu\text{as} = &-6.21^{+1.68}_{-1.66} ([G_{BP}-G_{RP}] - 1)^2 \nonumber \\ 
	& - 22.71^{+3.00}_{-2.96} ([G_{BP}-G_{RP}] - 1) - 27.35^{+1.49}_{-1.43}.
\end{align}
In considering separate quadratic parameterizations for sources $ G < 13 $ and $ G \geq 13 $, we find
\begin{align}
	\varpi_0 / \mu\text{as} = & -8.74^{+13.18}_{-13.42} ([G_{BP}-G_{RP}] - 1)^2 \quad\qquad (G < 13)\nonumber \\
	& - 85.67^{+13.25}_{-13.05} ([G_{BP}-G_{RP}] - 1) - 9.72^{+3.18}_{-3.06},
\end{align}
\begin{align}
	\varpi_0 / \mu\text{as} = & -13.58^{+1.88}_{-1.86} ([G_{BP}-G_{RP}] - 1)^2 \quad\qquad (G \geq 13)\nonumber \\
	& - 5.19^{+3.35}_{-3.52} ([G_{BP}-G_{RP}] - 1) - 31.82^{+1.59}_{-1.58}.
\end{align}
All three parameterizations are shown in \figurename~\ref{fig:ZPVariationsBPRP}, where the inferred functional forms are compared to individual estimates of $ \varpi_0 $ for each star using its measured parallax and the mean of its inferred distance posterior in the original $ K_s $ photometry model in \eqnname~\eqref{eq:TruePlx}. The separate parameterization for $ G < 13 $ appears not to be a good fit to the data; whereas, the parameterization for all $ G $ appears to behave better. This is further supported with the analysis of the variations with observed colour in bins, the results from which are also shown in \figurename~\ref{fig:ZPVariationsBPRP}.
We also find significant variation of the parallax zero point across the sky. Sky maps of zero points with hierarchical \texttt{HEALPIX} resolutions are shown in \figurename~\ref{fig:ZeroPointMap}. The maps appear to show variations with respect to Galactic latitude, which we attribute to the correlation of the characteristic scale length associated with the distance prior (\eqnname~\eqref{eq:DistancePrior}). The APOGEE data set reaches much deeper along the Galactic plane, meaning inferred values of $ L $ are larger close to the Galactic plane. The inferred $ \varpi_0 $ across each patch is then slightly correlated with $ L $. 

\begin{figure*}
	\centering
	\includegraphics[width=0.8\hsize]{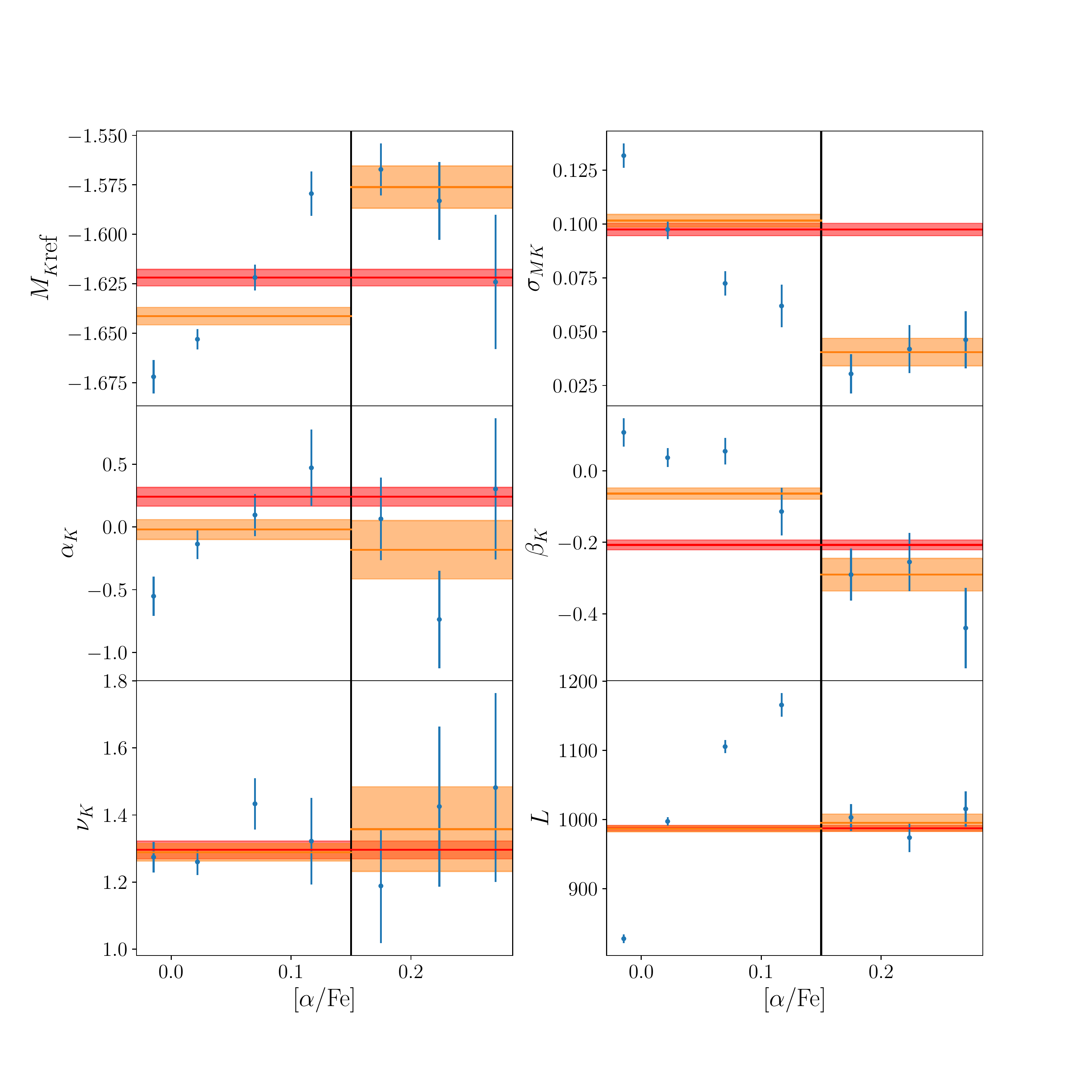}
	\caption{The dependence of red clump luminosity calibration parameters with $ [\alpha/\text{Fe}] $. \textit{Blue:} Median and $ \pm 1 \sigma $ error bars for posteriors in each $ [\alpha/\text{Fe}] $ bin. \textit{Red:} Median and $ \pm 1\sigma $ ranges for posterior distributions from the base $ K_s $ model for all $ G $. \textit{Orange:} Median and $ \pm 1\sigma $ ranges for posterior distributions for simply splitting between the low- and high-$ [\alpha/\text{Fe}] $ sub-populations. \textit{Black:} Split at $ [\alpha/\text{Fe}] = 0.15 $ dex between the low- and high-$ [\alpha/\text{Fe}] $ sub-populations.\label{fig:aFeBins}}
\end{figure*}

It is highly unlikely that the parallax zero point offset would have a pattern that follows Galactic latitude, as the \gaia\ satellite does not a priori know about the Galactic plane. The observed variation with sky position is therefore most likely due to intrinsic differences between the red clump stars at different Galactic latitudes that are not fully captured by our model. But the observed zero point variations across the sky still serve as an upper limit on the true variation of the zero point across the sky, because a large variation would be picked up by our model. In an analysis of the variations of $ \varpi_0 $ across the sky for all $ G $, we find the median parallax zero point (with $ \pm 1 \sigma $ dispersion) across patches to be $ \tilde{\varpi}_0 = -42.18^{+13.10}_{-14.96}~\mu\text{as} $. Similarly we find the spreads in patches for each split in $ G $ to be $ \tilde{\varpi}_0(G < 13) = -39.92^{+17.82}_{-24.96}~\mu\text{as} $ and $ \tilde{\varpi}_0(G \geq 0) = -41.03^{+13.11}_{-14.80}~\mu\text{as} $. We do not expect the \gaia\ parallax zero point to fluctuate by more than a few 10s of $ \mu\text{as} $ for any given observed magnitude and colour, as such fluctuations would be the ones captured by the inference rather than those induced by the degeneracies with $ L $.

\subsection{Modeling the Red Clump}\label{sec:RCModelingResults}

The analysis of the red clump luminosity calibration in $ K_s $ as a function of $ [\alpha/\text{Fe}] $ are shown in \figurename~\ref{fig:aFeBins}. The absolute magnitude exhibits clear trends for the low-alpha ($[\alpha/\text{Fe}] < 0.15\,\text{dex}$) and high-alpha ($[\alpha/\text{Fe}] > 0.15\,\text{dex}$) populations. In addition, the inverse variance averages of the calibration scatter in each population are FWHM$ /2.355 = 0.103 \pm 0.004 $ mag for $ [\alpha/\text{Fe}] < 0.15 $ dex, and FWHM$ /2.355= 0.040 \pm 0.008$ mag for $ [\alpha/\text{Fe}] \geq 0.15 $. While the luminosity dependence on [Fe/H], $ \beta_K $, appears to become stronger with $ [\alpha/\text{Fe}] $, the evolution of the $ (J_0 - K_0) $ slope $ \alpha_K $ does not seem to be as well constrained. This is understandable, as $ \alpha_K $ has also been less constrained than $ \beta_K $ for every other analysis. Nevertheless, this is further evidence for different behaviours between at least two sub-populations of red clump stars, and suggests that more detailed modeling of these populations may be necessary for precise use of red clump stars in the future. The small luminosity scatter for high-$ [\alpha/\text{Fe}] $ stars means that highly precise red-clump distances can be obtained for them.

Finally, we discuss the results of the verification of stellar models used in the APOGEE red clump catalogue discussed in \S\ref{sec:APOGEEVerifyMethod}. We find that the inferred residual $ M_\text{ref} = -0.036 \pm 0.004 $ with a dispersion of FWHM$ /2.355 = 0.096 \pm 0.004 $ shows the stellar evolution models describe the overall luminosity of the red clump quite well. However, we infer significant non-zero values of $ \alpha = 1.48 \pm 0.07 $ and $ \beta = -0.39 \pm 0.01 $ mag/dex, implying that the stellar evolution models leave significant residual luminosity dependencies on stellar temperature and metallicity [Fe/H].

\section{Discussion}\label{sec:Discussion}

\subsection{Consistency of zero point calibrations}\label{sec:DiscussionConsistency}

Throughout most of our analyses, the \gaia\ parallax zero point is usually inferred to be within the range of $ \varpi_0 \approx -47 $ to $ -49~\mu\text{as} $ when modeled as a constant. There are two main exceptions. First, the model using $ G $ photometry only infers $ \varpi_0 \approx -38~\mu\text{as} $. We attribute this to an incomplete understanding of interstellar extinction in $ G $, and we resolved the inconsistency by supplementing the model with $ K_s $ photometry. The $ K_s $ measurements include robust measurements of extinction, and shift the inference with $ G $ photometry to match with the other models. The second inconsistency comes from inferring two constant zero points for sources dimmer and brighter than $ G = 13 $. As similar inconsistencies were seen with the modeling of the zero point variations with $ G $, this seemed to be caused by a degeneracy associated with modeling both the zero point dependence and the red clump luminosity calibration simultaneously with photometric information. By either fixing the red clump absolute magnitude to a previously inferred value including an informed prior based on a previous posterior, these models give inferred zero points consistent with the all other models.

Finally, we discuss the possibility of attributing the variations in $ \varpi_0 $ across both $ G $ and sky position to a correlation with the given extinction corrections (namely $ A_K $ for most of our models). Running the base model using $ K_s $ photometry with an additional sample cut for $ 0 < A_K < 0.1 $ results in an inferred $ \varpi_0 = -40 \pm 1~\mu\text{as} $ with a corresponding $ M_{K,\text{ref}} \approx -1.611 $. Conversely, the sample containing $ A_K \geq 0.1 $ returns $ \varpi_0 = -36 \pm 2~\mu\text{as} $ with a corresponding $ M_{K,\text{ref}} = -1.720 $. This seems to contradict what the inferred parallax zero points with respect to $ G $ may suggest, but they in fact support each other. Within our model, a good estimate of the parallax zero point requires: 1) A good calibration of the red clump luminosity using nearby stars, and 2) A large enough sample of stars far away such that $ \varpi_0 $ is significant relative to the star's parallax. We can see the effects of removing either of these pillars in our analyses. The zero point inference split between bright/faint stars uses all stars to calibrate the luminosity; therefore, it retains a similar $ M_{K,\text{ref}} $ as the original model. The parallax zero points inferred for $ G < 13 $ are generally more inconsistent because these brighter sources tend to be closer to us; whereas, the fainter sources tend to have parallaxes that are more significanly affected by $ \varpi_0 $. Similarly, the low extinction sample above retains a consistent luminosity calibration, but fails to estimate a consistent parallax zero point because the stars are mostly closer to us. The high extinction sample is naively expected to provide a good $ \varpi_0 $, but it fails because it is unable to infer $ M_{K,\text{ref}} $ well because it has few bright, nearby stars. In conclusion, in order to use our models to obtain an estimate of the \gaia\ parallax zero point to both high accuracy \textit{and} precision, we require a large sample of stars that contains populations both close by (to get a good handle on the red clump luminosity) as well as far away (more significantly affected by $ \varpi_0 $).

\subsection{Comparison to other zero point work}

\begin{figure}
	\centering
	\includegraphics[width=\hsize]{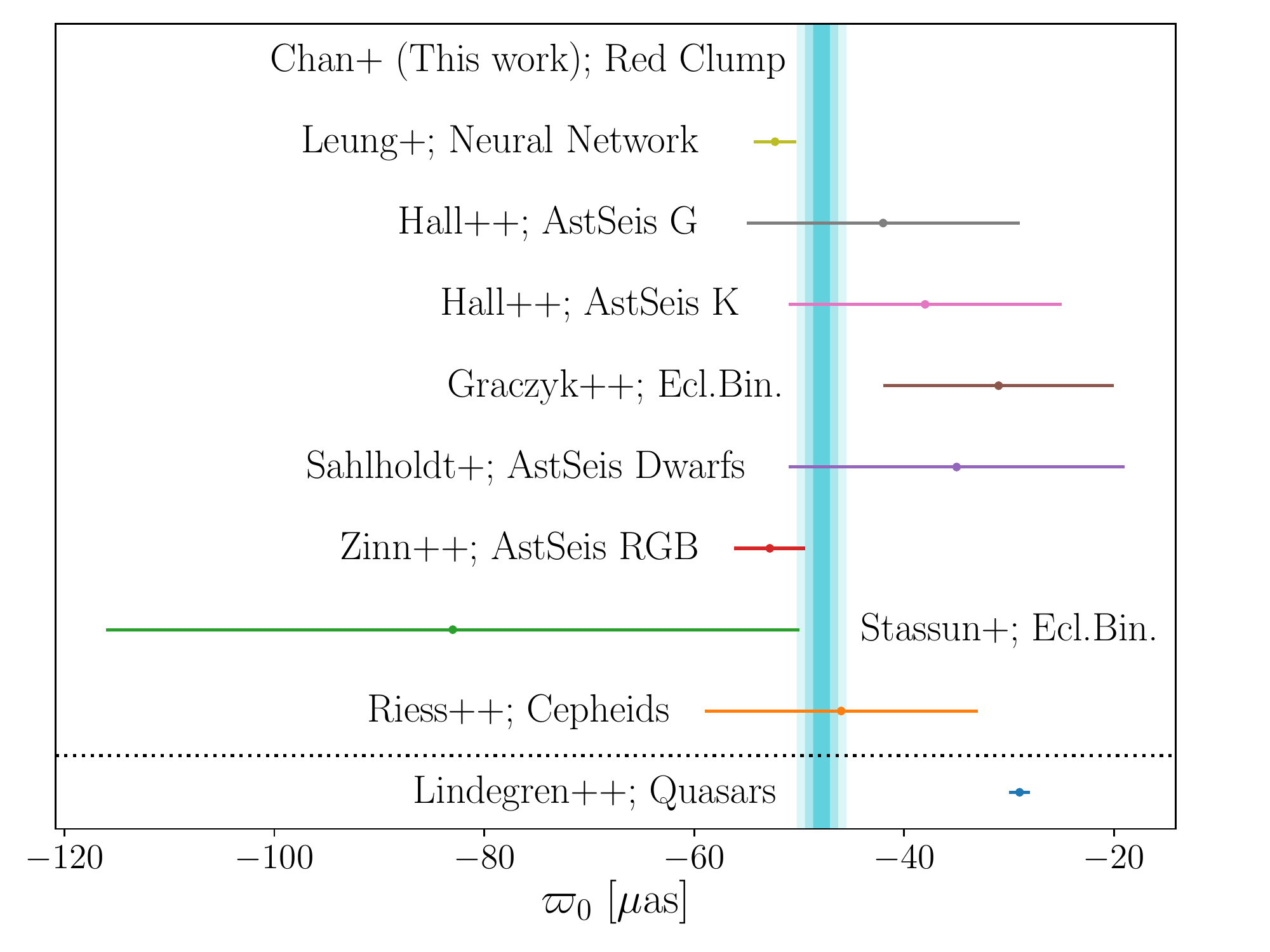}
	\caption{A comparison of reported \gaia\ parallax zero point values associated with Data Release 2. These measurements include those which make use of asteroseismology, neural networks applied to other red giants in APOGEE, eclipsing binaries, and Cepheid variables. The The \citet{AstSeisRGB} result is shown at $ \nu_\text{eff} = 1.5 $ and $ G = 12.2 $. The turquoise vertical stripes indicate the 1, 2, and 3$ \sigma $ regions of the \gaia\ DR2 parallax zero point that we report. Our measurement of the global \gaia\ zero point offset is consistent with all previous determinations using stars, and it is the most precise so far.\label{fig:ZPCompare}}
\end{figure}

With the resolution of tensions between the models presented in this paper, we choose to report a constant \gaia\ DR2 parallax zero point of $ \varpi_0 = -47.97 \pm 0.79~\mu\text{as} $ as inferred by our base model with $ K_s $ photometry. This measurement can be seen in \figurename~\ref{fig:ZPCompare} in comparison to other values reported in the literature discussed previously. Note that \citet{KhanAst} also finds a systematic offset in \gaia\ DR2 parallaxes in the range of $ -45 $ to $ -55~\mu\text{as} $. Our inferred parallax zero point is the most precise to date, and is in good agreement with most other measurements. Note that the red clump occupies a specific space in both observed colour ($ 1 \lesssim G_{BP}-G_{RP} \lesssim 3.5 $) and magnitude ($ 9.5 \lesssim G \lesssim 18 $), so the inferred zero point reported in this paper becomes less accurate outside of this space. This can explain why the \gaia\ zero point reported by \citet{GaiaDR2Astrometry} is quite discrepant with our own, as quasars typically occupy a dimmer and bluer part of the \gaia\ observation space.

Variations of the parallax zero point appear to be best described using models discussed in this paper with \eqnname~\eqref{eq:Gdepl13} and \eqnname~\eqref{eq:Gdepg13} for $ G $ dependent fluctuations as well as \eqnname~\eqref{eq:BPRPdepAllG} for observed colour dependent fluctuations. This is in contrast to the functional parameterizations of the dependences reported by \citet{AstroNNZP}. The data sets in both studies occupy similar regions of observed colour and magnitude, yet we infer variations that are much larger in amplitude. In addition, both quadratic dependences analyzed in this paper reflect parabolas with negative curvature; whereas, the parameterizations reported in \citet{AstroNNZP} exhibit positive curvature. We are uncertain as to why this may be the case.

Finally the variations of $ \varpi_0 $ with sky position are inferred here to fluctuate between $ -80~\mu\text{as} $ to $ -10~\mu\text{as} $, with a typical scatter across the sky of approximately $ 15~\mu\text{as} $ about $ -40~\mu\text{as} $. We conclude that the \gaia\ parallax zero point must not vary by more than a few 10s of $ \mu\text{as} $ across the sky. Our range of systematic fluctuations is similar to the $ \sim 100 \mu\text{as} $ peak-to-peak fluctuations reported by \citet{GaiaDR2CatValid}. It is also worth noting that \citet{KhanAst} also find peak-to-peak variations of the same order for the parallax zero point across the sky

\subsection{Comparison to previous calibrations of the RC}

We infer the absolute magnitude of the red clump in the 2MASS as well as \gaia\ photometry bands. The inferred absolute magnitudes in $ K_s $ and $ G $ appear to be consistent with those reported by \citet{HallAstSeis}. We remark that, while the inferred absolute magnitude in $ K_s $ are in agreement with that reported in \citet{HawkinsRC}, the absolute magnitudes we infer in $ J $, $ H $, and $ G $ are not compatible. Of course, this only applies to the estimate of the peak of the red clump luminosity distribution or the absolute magnitude of a typical red clump star. The scatters in absolute magnitude inferred in both this paper and in \citet{HawkinsRC} are large enough to contain both estimates. It is possible that incomplete modeling of the red clump sub-populations as described in \S\ref{sec:aFeMethod} could contribute to such differences in red clump calibrations.

\section{Conclusions}\label{sec:Conclusion}

The advent of the \gaia\ mission has provided the astronomical community with an excellent opportunity to use ultra precision astrometry; however, a lack of understanding of the parallax zero point has prevented us from unlocking the full potential of \gaia\ data. A precise measurement of the \gaia\ parallax zero point requires a large data set, as well as a robust understanding of the methods involved. We have presented several hierarchical probabilistic models using red clump stars to infer the parallax zero point, while simultaneously calibrating an empirical parameterization for the red clump luminosity. In doing so, we infer the \gaia\ DR2 parallax zero point to be $ \varpi_0 = -48 \pm 1~\mu\text{as} $ using $ K_s $ photometry. Models using other 2MASS or \gaia\ photometry allow for consistent estimates of the zero point. The use of a Student's $t$-distribution appears to describe the distribution of luminosities in our red clump sample quite well. We also report the absolute magnitude of the red clump to be $ M_\text{ref} = -1.622 \pm 0.004 $ in $ K_s $, $ M_\text{ref} = 0.435 \pm 0.004 $ in $ G $, $ M_\text{ref} = -1.019 \pm 0.004 $ in $ J $, and $ M_\text{ref} = -1.516 \pm 0.004 $ in $ H $. We find the intrinsic spread of the red clump to be $ \sim 0.09 $ in $ J $, $ H $, and $ K_s $. The scatter in $ G $ is $ \sim 0.12 $, and this larger value can be attributed to an incomplete understanding of interstellar extinction. Each probabilistic model also infers the distance to every star used as input, yielding typical distance estimates of $ \sim 10\% $.

Additions to the base probabilistic model allow for more detailed investigation into either into the variations of the \gaia\ parallax zero point or the red clump luminosity calibration. We find that the variations in the zero point are most significant for dim sources $ G \gtrsim 16 $, while the zero point offset is constant at brighter magnitudes. Note that less than $ 7\% $ of our total sample lies within $ G \gtrsim 16 $, so the inferred parallax systematics may not be as accurate as in brighter bins. The dependence of the zero point on observed colour is can also be parameterized with a quadratic form. Fluctuations of the zero point across the sky are difficult to infer, but we limit them to be less than a few 10s of $ \mu\text{as} $. We also find significant variations of the red-clump luminosity model with $[\alpha/\text{Fe}]$ and in particular with different trends across low- and high-$ \alpha/\text{Fe} $ sub-populations.

The sheer size of the red clump sample has allowed us to estimate the \gaia\ DR2 parallax zero point to approximately $ 1.6\% $. This is the highest precision estimate of the parallax zero point to date, and presents the community with a fantastic outlook for using \gaia\ for high-precision distance estimates. Although it is expected to be smaller in amplitude and better understood, the parallax zero point will still be present in future \gaia\ data releases. We also expect the quality of \gaia\ data to be better in the future, implying that these probabilistic methods will eventually allow us to infer distances to stars to high enough accuracy and precision to be impactful in areas of galactic dynamics, constructing local distance ladders, and much more.

\section*{Acknowledgements}

We thank the anonymous referee, Anthony Brown, Fr\'ed\'eric Arenou, and Beno\^it Mosser for thoughtful feedback, suggestions, and discussions.

VCC received support from the Ontario Graduate Scholarship/Queen Elizabeth II/ Walter John Helm Graduate Scholarships In Science And Technology.

VCC and JB received support from the Natural Sciences and Engineering Research Council of Canada (NSERC; funding reference number RGPIN-2015-05235) and from an Ontario Early Researcher Award (ER16-12-061). JB also received partial support from an Alfred P. Sloan Fellowship.

This work presents results from the European Space Agency (ESA) space mission Gaia. Gaia data are being processed by the Gaia Data Processing and Analysis Consortium (DPAC). Funding for the DPAC is provided by national institutions, in particular the institutions participating in the Gaia MultiLateral Agreement (MLA).

Funding for the Sloan Digital Sky Survey IV has been provided by the Alfred P. Sloan Foundation, the U.S. Department of Energy Office of Science, and the Participating Institutions.



\bibliographystyle{mnras}
\interlinepenalty=10000
\bibliography{redclumpbib}

\bsp	
\label{lastpage}
\end{document}